\documentclass[useAMS,usenatbib]{mn2e}
\usepackage{graphicx}  
\usepackage{epstopdf}                %
\usepackage{float}
\usepackage{placeins}
\usepackage{longtable}
\usepackage{supertabular}
\usepackage{ulem}
\usepackage[usenames,dvipsnames]{xcolor}
\usepackage[flushleft]{threeparttable}

 \title[Investigation of open clusters based on IPHAS  and APASS survey data]
 {Investigation of open clusters based on IPHAS  and APASS survey data}
 \author[]
 {
 A. K. Dambis$^{1}$\thanks{E-mail: dambis@yandex.ru}, E. V. Glushkova$^{1,2}$, 
 L. N. Berdnikov$^{1,3}$, Y. C. Joshi$^{4}$
 \newauthor A. K. Pandey$^{4}$ \\
 $^{1}$Sternberg Astronomical Institute, Lomonosov Moscow State University, 
13, Universitetskii prospect, Moscow, 119992, Russia\\
 $^{2}$Physics Faculty, Lomonosov Moscow State University, 1, bld.2, Leninskie Gory, Moscow,  119991, Russia\\
 $^{3}$Astronomy and Astrophysics Research division, Entoto Observatory and Research Center, P.O.Box 8412, Addis Ababa, Ethiopia   \\
$^{4}$Aryabhatta Research Institute of Observational Sciences, Nainital 263002, India\\
 }

\begin{document}
 \date{Accepted 2016 MONTH XX.} %%
 \pagerange{\pageref{firstpage}--\pageref{lastpage}} \pubyear{2016}
 \maketitle
 \label{firstpage}

\begin{abstract}
We adapt the classical $Q$-method based on a reddening-free parameter constructed from three
passband magnitudes to the filter set of IPHAS survey and combine it with
the maximum-likelihood-based cluster parameter estimator by \citet{naylor_jeffries} to 
determine the extinction,  heliocentric distances, and ages of young 
open clusters using H$_{\alpha}$ri data. The method is also adapted for the case
of significant variations of extinction across the cluster field.
Our technique is validated by comparing the colour excesses, 
distances, and ages determined in this study with the most bona fide values reported for the 18 well-studied
young open clusters in the past and a fairly good agreement is found between our 
extinction and distance estimates and earlier published 
results, although our age estimates are not very consistent with those published by other authors. We also show that individual extinction values can be 
determined rather accurately for stars with $(r-i)>$~0.1.
Our results open up a prospect for determining a uniform  set of parameters for northern clusters based 
on homogeneous photometric data,  and for searching for new, hitherto undiscovered open clusters.     
\end{abstract}


 \section{INTRODUCTION}
Galactic clusters have been popular among the researchers for several centuries, particularly open
clusters (OC), because they are the ``building blocks" that make up our 
Milky Way Galaxy. Large-scale sky surveys have brought about the discovery of 
many hitherto unknown open clusters, which soon outnumbered those already known 
over the past 150 years. Currently, about 4000 open clusters are known and the 
cluster sample within about 1.7--1.8 kpc from the Sun can be considered to be almost complete
\citep{joshi, joshi16}. 
However, a homogeneous catalogue of fundamental open-cluster parameters, primarily 
colour excesses, ages, and heliocentric distances, is needed for such a rich subsystem 
to be used for the study of the structure, kinematics, and evolution of the Galactic 
disk or for solving the problems of star formation. There are ongoing attempts to 
create such a catalogue. Some authors try to produce the most extensive possible catalogue 
of cluster parameters by compiling published data \citep{dias02} or by 
determining the cluster parameters based on published observational data 
\citep{kharchenko}, however, the reported parameters in these compilations seem to be insufficiently 
accurate for the study of the properties of the Galactic disk \citep{netopil}. 
Other researchers determined the cluster parameters from their own observational data, 
however, such catalogues are rather small and may take several decades to complete 
\citep{bragaglia, sung, caetano}. The main 
problem with the determination of fundamental cluster parameters like age, distance, 
and colour excess – is that an empirical ZAMS or isochrone cannot always be unambiguously 
fitted to the cluster main sequence on colour-magnitude diagrams. Despite the 
development of new techniques and algorithms for the determination of the basic 
parameters of open clusters 
\citep{naylor, dias12, popescu, perren}, 
no breakthrough has been achieved so far to prepare a uniform and accurate 
catalogue of cluster parameters. The most bona fide estimates of colour excess $E_{B-V}$ 
can be obtained by fitting ZAMS to the dwarf sequence on the (U-B, B-V) colour-colour 
diagram. However, the intrinsic colour line for late-type stars in colour-colour diagrams 
runs almost parallel to the reddening line, therefore
this method is 
applicable only to relatively young clusters. Since most of the large 
surveys lack U-band photometry, primarily because it requires long signal integration 
times, hence reddening cannot be determined. 
However, we show below that there is another kind of colour-colour diagrams 
based on already available observational data that can be used to unambiguously and 
rather accurately determine the colour excesses of young clusters. In the present
study we use $r$, $i$, and H$_{\alpha}$-band magnitudes from IPHAS DR2 survey \citep{barentsen, drew}
to determine colour excesses of some poorly investigated clusters.  
%\newpage

\section{DATA}
\label{data}
As we already pointed out above,
our source of photometry is the INT Photometric H$_{\alpha}$ Survey of the Northern Galactic Plane
(IPHAS), which is based on observations made using wide field camera
(WFC) on the 2.5m Isaac Newton Telescope (INT) with
Sloan $r$- and $i$-band filters and H$_{\alpha}$ narrow-band filter 
\citep{drew, gonzales_solares}. The photometric data for cluster stars used in
this study were adopted from the latest Data release 2 of the survey catalogue, which
covers the entire northern Galactic plane and gives $r$, $i$, and H$_{\alpha}$– band magnitudes for about 
219 million sources covering Galactic longitudes l = 30$^o$ - 215$^o$ 
and latitudes $|b|< 5^o$
down to a limiting magnitude of 21.2~mag, 20.0~mag, and 20.3~mag in the r, i, and 
H$_{\alpha}$ filters, respectively.
\citep{barentsen}.

\section{METHOD}
\label{method}

\subsection{Basic ideas}

\citet{sale} used IPHAS photometry to construct a 3D extinction map in the 
northern part of the Galactic plane. Their procedure was based on the fact that (r - H$_{\alpha}$) 
colour index depends primarily on the equivalent width of the H$_{\alpha}$ line
hence on the effective temperature 
of the source. Their extinction map has an angular resolution 
and distance sampling of 10~arcmin and 100~pc, respectively, 
and gives the distribution of monochromatic extinction A0 at 5495\AA. \citet{sale} 
constructed their map using the data for 38 million stars for each of which they determined the 
distance, reddening, mass, surface gravity, and effective temperature. However, according to the 
above authors, the calculation of all these parameters are subjected to a set of priors 
as well as on a survey selection function and should be used with much caution and only as approximate 
estimates of the real parameters. It is also important that \citet{sale} 
actually treat the overall extinction distribution (and hence the extinction distribution in open
clusters) as somewhat smoothed, albeit hierarchical, 
with a limited sampling mentioned above (10~arcmin and 100~pc, respectively) and ignore the 
existence of compact groups of stars with small spatial and age dispersion. 
The technique proposed in this paper, 
on the contrary, is designed especially for open clusters and uses their 
compact size in the physical space and in the space of ages and interstellar-extinction 
(we first assume it to be sufficiently uniform for all cluster members and defer the analysis of the 
variable-extinction case to Section~3.2).

We use somewhat modified colour-colour diagrams based on the IPHAS data. Our idea is to adapt the
classical $Q$-method \citep{qmethod} and construct
a reddening-free photometric index based on the $r$, $i$, and H$_{\alpha}$
magnitudes provided by the survey. 
Given three different passbands we have two independent colour 
indices (e.g., $r$-H$_{\alpha}$ and $r$-$i$) and only one possible reddening-free index 
(up to an arbitrary linear transformation) linearly depending on the survey magnitudes:
H$_{\alpha}$~index = ($r$-$H_{\alpha}$)~-~($E_{r-H\alpha}$/$E_{r-i}$)($r-i$) = 
($r$-$H_{\alpha}$)~-0.245($r-i$) = 0.755$r$~+~0.245$i$~–-~$H_{\alpha}$, where
we adopt ($E_{r-H\alpha}$/$E_{r-i}$) = 0.245 in accordance with the \cite{cardelli} reddening law. 
The resulting index  has 
simple physical meaning: linear combination 0.755$r$~+~0.245$i$ 
interpolates between $r$- and $i$-band magnitudes and sort of simulates a 
very broad band $H_{\alpha}$ magnitude, and hence 0.755$r$~+~0.245$i$~–-~$H_{\alpha}$ serves as an 
extinction-independent measure of the $H_{\alpha}$ line strength (with the reverse sign) 
similar to the $\beta$ index of Stroemgren system, which
characterises the strength of the H$_{\beta}$ absorption line. Hydrogen line strengths are 
known to be the highest in stars of spectral type A2 \citep{mcbride} 
and hence H$_{\alpha}$~index has a minimum at the corresponding color index.
We therefore use  the (0.755r + 0.245i - H$_{\alpha}$, $r$-$i$) diagram
as our chief extinction-measuring tool in this study. By construction, the reddening lines in the 
resulting diagrams 
are parallel to the horizontal axis. 

\begin{figure*}
\includegraphics[width=\linewidth]{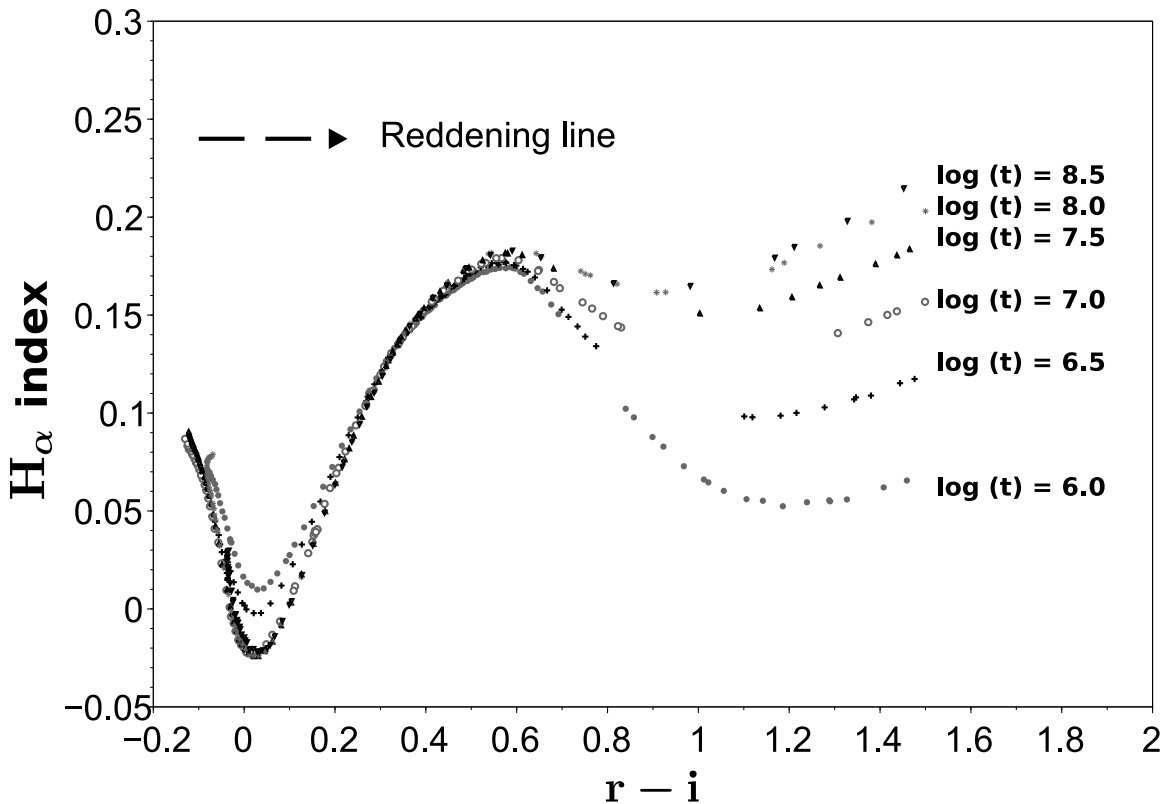}
\caption{Padova isochrones on the ($r-i$,H$_{\alpha}$~index) colour-colour diagram.}
\label{cciso}
\end{figure*}

As expected, the corresponding most recent ($r-i$,H$_{\alpha}$~index) theoretical isochrones constructed in the Padova system 
\citep{chen1, chen2, tang} all have a minimum at the same colour excess ($r-i$) for all ages spanning from 
log~($t$) = 6.0 to log~($t$) = 8.5.  Fig.~\ref{cciso} shows ($r-i$,H$_{\alpha}$~index) colour-colour diagram for
the solar-metallicity isochrones with ages in the above 
interval. A reddening line is shown in the top left corner. Note that the minima of all isochrones are 
located at the same horizontal coordinate, but are somewhat shifted in the vertical direction. 
Table 1 lists the $(r-i)_{min}$ values for solar-metallicity models as well as for models with 
[Fe/H]=-0.5 and [Fe/H]=+0.5 (note that in the set of recent Padova stellar
evolution models used here the solar distribution of heavy elements is adopted 
from the by \citet{caffau}, corresponding to a Sun’s metallicity $Z$~=~0.0152). As is evident from the table, the abscissa of the minimum 
appears to be highly invariant not only over a broad age interval, but also highly stable 
against metallicity variations: the mean $(r-i)_{min}$ value is equal to +0.026~mag with a 
standard deviation of 0.004 (per isochrone) and the maximum deviation does not exceed 0.006. 
The fact that for ($r-i$) lower than 0.5 metallicity and age have practically no 
effect on the observed colors is well known, however, we would nevertheless like to emphasize this
point quantitatively to show the robustness of our extinction estimates and demonstrate
that the position of the minimum of the H$_{\alpha}$~index~vs~($r-i$) curve is an excellent reddening 
indicator for clusters that is practically insensitive to cluster age and metallicity. Older 
isochrones have no left wing (the part of the curve with intrinsic colour index $(r-i)_0 < 0.0$) 
and that is why we do not show them here. This is because, as we already pointed out above, 
H$_{\alpha}$ absorption line reaches maximum strength in the spectra of A2-type stars, 
which are absent in the main sequences of clusters older than log~($t$)~=~8.5.

\begin{figure}
\includegraphics[width=\linewidth]{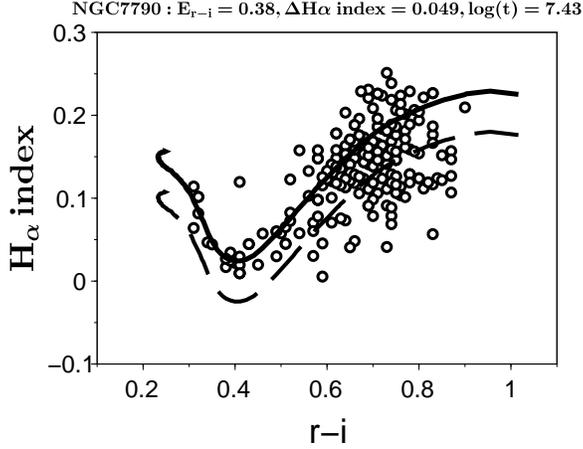}
\caption{Colour-colour diagram for stars within 3~arcmin of the NGC 7790 cluster centre and the 
superimposed log~($t$)=7.2 isochrone shifted by 0.38~mag along the horizontal axis (the thin dashed curve) and the
same isochrone shifted by 0.38~mag along the horizontal axis and by $\Delta$~H$_{\alpha}$~index~=~+0.049 along the vertical axis (the thick solid curve).}
\label{cc7790}
\end{figure}

Hence if a cluster contains enough early A-type main-sequence stars and extinction variations across
it are not significant enough (the case of highly variable reddening is addressed in Section~3.2), 
its average $E_{r-i}$ colour excess can theoretically
be easily determined by fitting the cluster MS with an isochrone of any age in the log~($t$)~=~6.0~–-~8.5 
interval simply by shifting the latter along the horizontal axis. We illustrate this in Fig.~\ref{cc7790}, 
where we show the ($r-i$,H$_{\alpha}$~index) diagram for the cluster NGC 7790 with the superimposed log~($t$)~=7.2 
isochrone shifted horizontally by 0.38~mag. Note, however, that in practice 
a vertical offset (in the H$_{\alpha}$~index) has also to be applied because, although the shape
of the observed diagram matches fairly well the theoretical curve, the observed H$_{\alpha}$~index
values are located above the theoretical line, which runs about 0.049 below the lower envelope of the observed
points. As we show below, this is a common behaviour
and the vertical offsets of the lower envelope are more or less the same for all clusters. This offset 
is most likely due to some systematic zero point mismatches between the color indices as
represented by the actual IPHAS catalogue and the corresponding theoretically computed colors.

Now, having determined the colour excess $E_{r-i}$ of a cluster, we can proceed with 
estimating such fundamental parameters as the distance and age. To this end, we use
the ($r-i$)--$r$ diagram and fit an appropriate theoretical isochrone by shifting it 
horizontally by the already known $E_{r-i}$ and vertically by the unknown apparent
$r$-band distance modulus, $DM_r$=$r-M_r$. The true distance modulus is then
determined as $DM_0$=$DM_r$--3.98$E_{r-i}$ and the distance in kpc as 
$d$=10$^{0.2(DM_0-10)}$. Note, however, that whereas the cluster distance can be
inferred rather accurately, IPHAS data does not allow age to be determined reliably
enough for most of the clusters because stars brighter than 
$r$ = 13~mag, $i$=12~mag are saturated in the IPHAS survey 
and their magnitudes are subject to systematic errors \citep{barentsen}. As a result, the MS turnoff, which is the region where age
differences are  most conspicuous, becomes unavailable.
To partially mitigate this effect,
we use $ri$ photometry of the APASS survey \citep{apass1, apass2} for such stars, which we transform
to the IPHAS photometric systems via the following equations \citep{barentsen}:
\begin{equation} 
r_{\rm IPHAS} = r_{\rm APASS} - 0.121 + 0.032(r-i)_{\rm APASS} 
\label{apass_r} 
\end{equation}

\begin{equation}
i_{\rm IPHAS} = i_{\rm APASS} - 0.364 + 0.006(r-i)_{\rm APASS} 
\label{apass_i}
\end{equation}
 
\begin{figure}
\includegraphics[width=\linewidth]{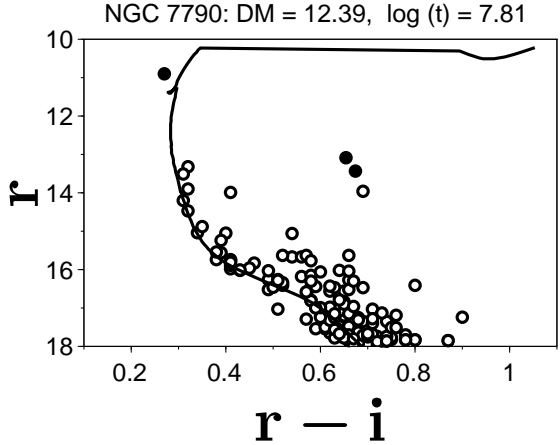}
\caption{The ($r-i$, $r$) colour-magnitude diagram for stars within 3~arcmin from the centre of the 
NGC 7790 cluster with the superimposed log~($t$)=7.81 isochrone shifted by 0.38 along the
horizontal axis and by 13.90~mag along the vertical axis (the solid curve).
The open circles represent unsaturated stars with IPHAS data and 
the filled circles, stars with APASS $ri$ photometry converted to the IPHAS system
in accordance with Eqs.~\ref{apass_r} and \ref{apass_i}. Given the above colour excess
estimate $E_{r-i}$=0.38~mag the true distance modulus is $(m-M)_0$~=12.39, which
corresponds to the distance of $d$=3006~pc.}
\label{cmd7790_rir}
\end{figure}

In this way we push the saturation limit brightwards by $\sim$~3~mag.
We illustrate this in Fig.~\ref{cmd7790_rir}, 
where we show the ($r-i$)--$r$ diagram for the cluster NGC 7790 with the
superimposed log~($t$)=7.81 
isochrone shifted horizontally by $E_{r-i}$=0.38~mag and vertically by $DM_r$=13.90.  The resulting true distance modulus estimate of the cluster is 
$DM_0$~=~$DM_r$--3.98$E_{r-i}$~=11.800, which corresponds to the distance of 3006~pc.

\begin{figure}
\includegraphics[width=1.1\linewidth]{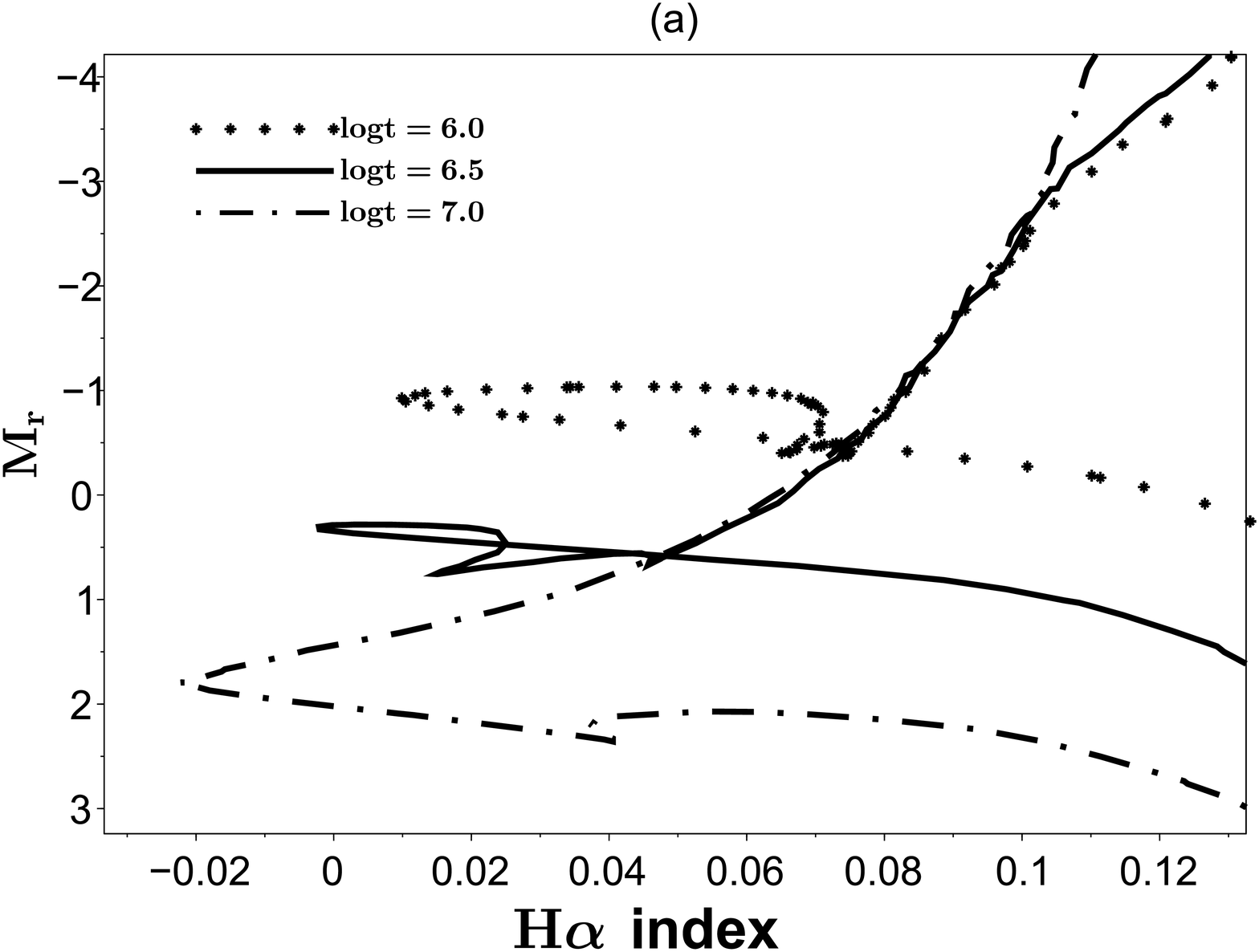}
\includegraphics[width=1.1\linewidth]{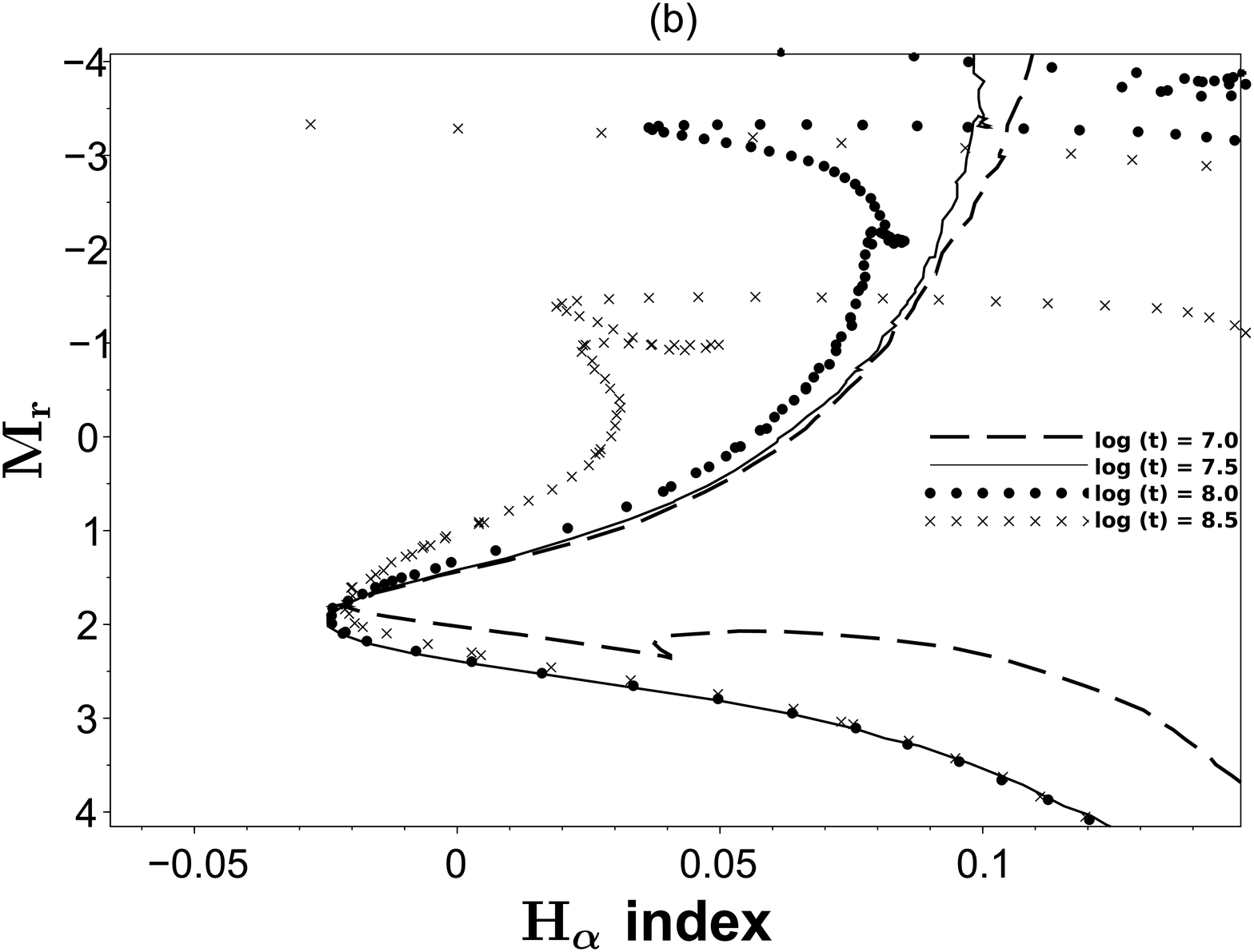}
\caption{The (H$_{\alpha}$~index, $M_r$) diagram for isochrones: (a) log~($t$)~=6.0, 6.5,  7.0 
and (b) log~($t$)~=7.0, 7.5,  8.0, and 8.5.}
\label{cmdiso}
\end{figure}

\begin{figure}
\includegraphics[width=\linewidth]{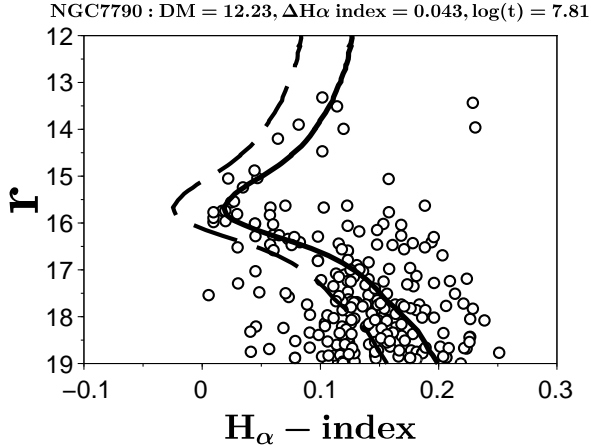}
\caption{The (H$_{\alpha}$~index, $r$) ``colour-magnitude" diagram for stars within 
3~arcmin from the centre of the 
NGC~7790 cluster with the superimposed log~($t$)=7.81 isochrone shifted by 13.74~mag along the vertical axis (the dashed curve)
and the same isochrone shifted by 13.74~mag 
along the vertical axis and by $\Delta$~H$_{\alpha}$~index~=~+0.043 along the horizontal axis (the solid curve). Given the above colour excess
estimate $E_{r-i}$=0.38~mag the true distance modulus is $(m-M)_0$~=12.23, which
corresponds to the distance of $d$=2793~pc.}
\label{cmd7790_hr}
\end{figure}

The (H$_{\alpha}$~index, $M_r$) isochrones with ages no greater than log~($t$)=8.5 also have an interesting feature: 
a characteristic leftward tip. 
Fig.~\ref{cmdiso} shows the theoretical isochrones log~($t$)= 6.0, 6.5, 7.0, 7.5, 8.0, and 8.5 
in the $M_r$~vs~H$_{\alpha}$~index plane. The tip locations of all isochrones with ages log~($t$)~=7.0 or older 
have practically the same horizontal and, what is important, vertical  
coordinate. The apparent distance modulus $(m-M)_r$ of a comparatively young cluster can be determined by shifting 
the cluster main sequence on the (H$_{\alpha}$~index, $M_r$) diagram to fit the isochrone (note that it should
be accompanied by a shift in the H$_{\alpha}$~index to correct for a small inconsistency between the theoretical and observed 
colors). The distance to the cluster 
can then be determined from this distance modulus and the colour excess earlier determined from the 
($r-i$,H$_{\alpha}$~index) diagram. Fig.~\ref{cmd7790_hr} shows the (H$_{\alpha}$~index, $r$) ``colour-magnitude" diagram 
for the cluster NGC~7790 superimposed with the log~($t$)=7.81 isochrone shifted by 13.74~mag 
along the vertical 
axis and by $\Delta$~H$_{\alpha}$~index~=~+0.043 along the horizontal axis. 
Thus the distance to the cluster is equal to 2793~pc, which more or less agrees with the
2944 pc estimate listed in the catalogue by \citep{dias02} although is appreciably
shorter than the 3548~pc estimate by \citet{phelps}.

\subsection{The case of non-uniform extinction}
\label{nonuniform}

\begin{figure}
\includegraphics[width=1.0\linewidth]{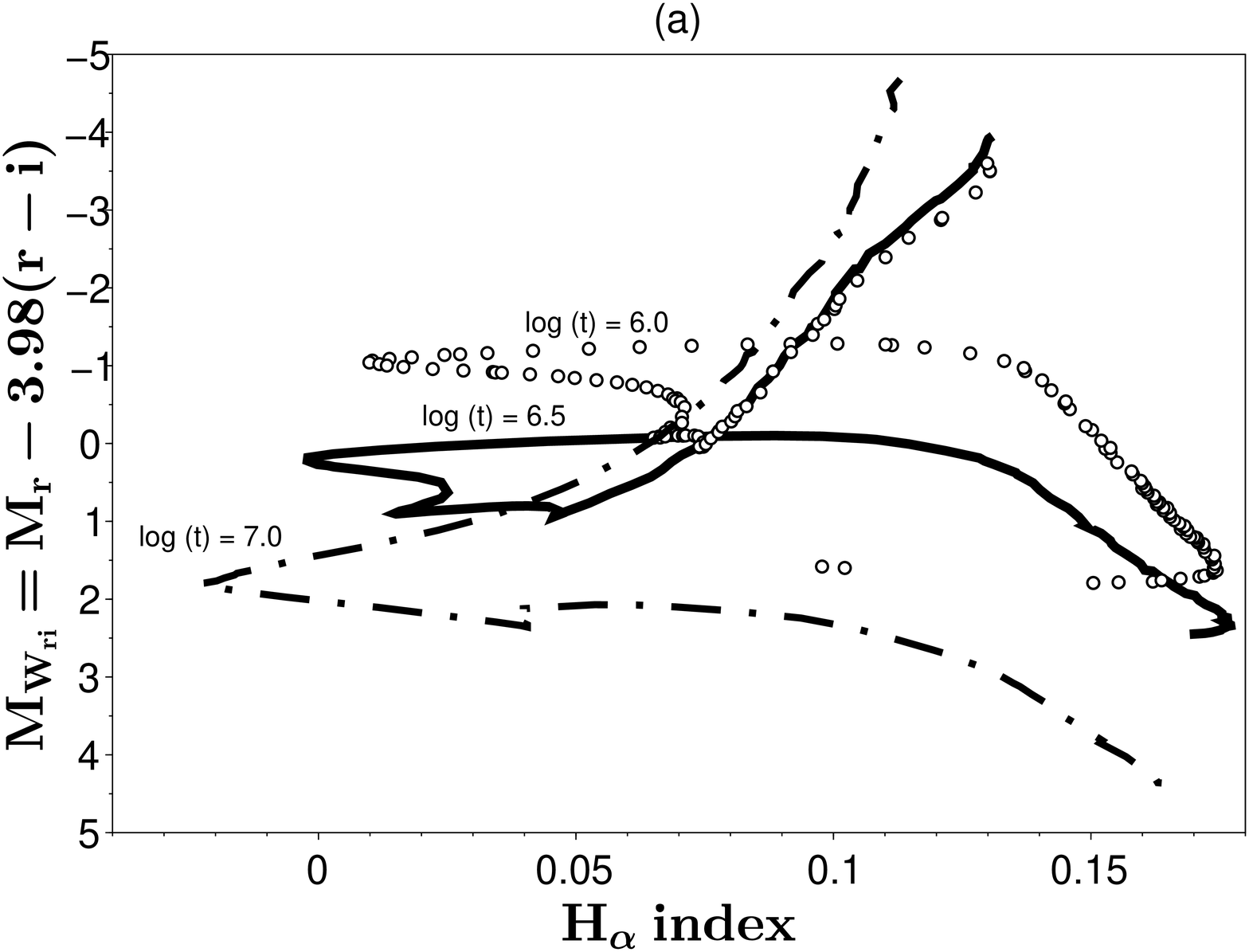}
\includegraphics[width=1.0\linewidth]{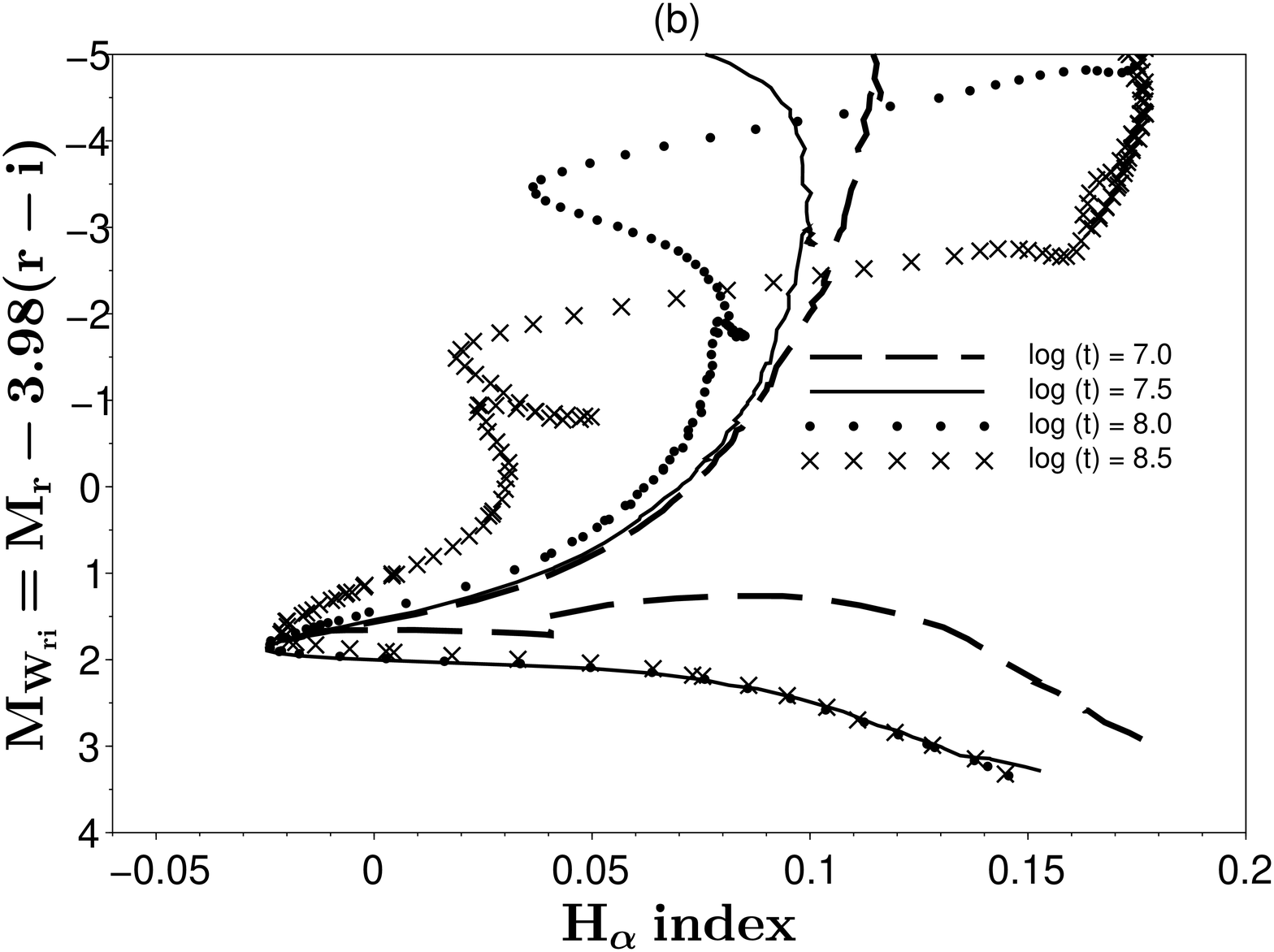}
\caption{The (H$_{\alpha}$~index, $M_{W_{ri}}$) diagram for isochrones: (a) log~($t$)~=6.0, 6.5 
and (b) log~($t$)~=7.0, 7.5,  8.0, and 8.5. }
\label{cmdhw}
\end{figure}

\begin{figure}
\includegraphics[width=0.9\linewidth]{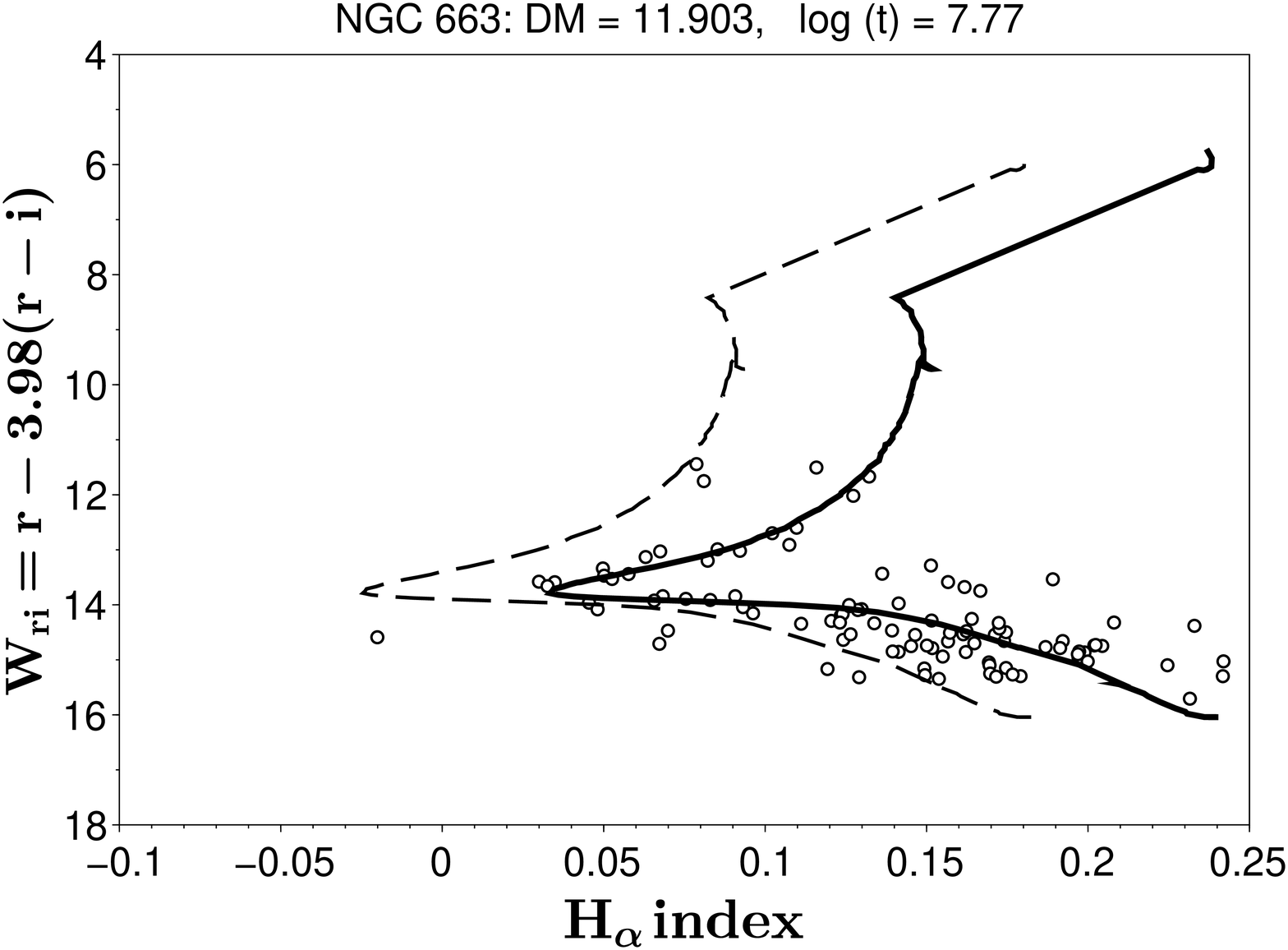}
\includegraphics[width=0.9\linewidth]{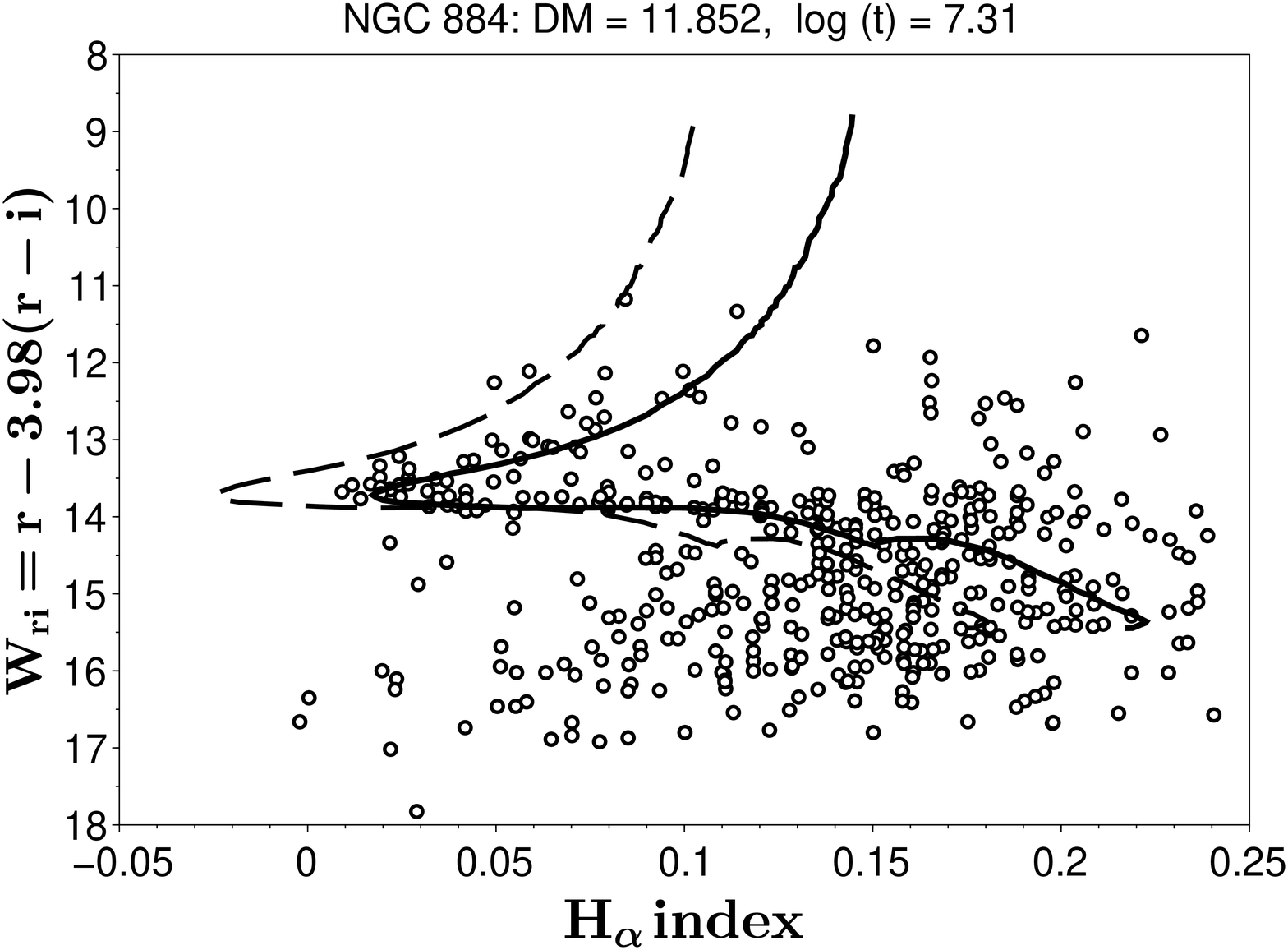}

\caption{Top panel: the (H$_{\alpha}$~index,$W_{ri}$) ``colour-magnitude" diagram for stars within 3~arcmin and from the centre of the 
NGC 663 cluster with the superimposed log~($t$)=7.77 isochrone shifted by 11.903~mag along the vertical axis (the dashed curve)
and the same isochrone shifted by 11.903~mag (the true distance modulus)
along the vertical axis and by $\Delta$~H$_{\alpha}$~index~=~+0.058 along the horizontal axis (the solid curve). 
Bottom panel: the (H$_{\alpha}$~index, $W_{ri}$) ``colour-magnitude" diagram for stars within 5~arcmin and from the centre of the 
NGC 884 cluster with the superimposed log~($t$)=7.31 isochrone shifted by 11.852~mag along the vertical axis (the dashed curve)
and the same isochrone shifted by 11.852~mag (the true distance modulus)
along the vertical axis and by $\Delta$~H$_{\alpha}$~index~=~+0.042 along the horizontal axis (the solid curve).  }
\label{cmdreal_hw}
\end{figure}

So far, we assumed that reddening is more or less uniform across the cluster and used
the ($r-i$,H$_{\alpha}$~index) diagram to determine it. This, however, is by no means 
always the case, and we therefore have to adapt our method to make it applicable to clusters 
with variable extinction. Our solution is to operate 
with extinction-independent quantities exclusively, and our tool of choice in this case
is the (H$_{\alpha}$~index, $W_{ri}$) diagram, where $W_{ri}$=$r-3.98(r-i)$ is  extinction-independent Wesenheit index. Fig.~\ref{cmdhw} shows the theoretical isochrones 
for log~($t$)= 6.0, 6.5 (the two youngest isochrones in the top panel) and log~($t$)=  7.0, 7.5, 8.0, and 8.5  (the bottom panel)
in the $M_{W_{ri}}$~vs~H$_{\alpha}$~index plane (here $M_{W_{ri}}$ is the
absolute Wesenheit magnitude). Note again the characteristic
leftward tip where the two branches of the isochrone meet. This  time the branch
corresponding to lower-mass stars  remains almost flat from H$_{\alpha}$~index~$\sim$~0 to
H$_{\alpha}$~index~$\sim$~0.1 and then goes down toward fainter absolute Wesenheit
magnitides. Furthermore, the absolute Wesenheit magnitude $M_{W_{ri}}$ in the flat part
of the diagram is practically independent of age for  log~($t$)~$\geq$~7.5.
Hence some knowledge of the cluster age (if log~($t$)~$\sim$~6.0--7.5) 
or just the knowledge that it is not younger than log~($t$)~$\leq$~7.5 is sufficient for
estimating rather accurately the cluster distance. Hereafter we fit the observed 
(H$_{\alpha}$~index, $W_{ri}$) diagram to a theoretical diagram to simultaneously
determine the cluster age and distance. Like in the case of the (H$_{\alpha}$~index, $M_r$) 
diagram we also introduce an extra free parameter to 
allow for a shift  in the H$_{\alpha}$~index 
($\Delta$~H$_{\alpha}$~index) in the sense 
H$_{\alpha}$~index (observed)~=~H$_{\alpha}$~index (theoretical)~+~$\Delta$~H$_{\alpha}$~index to correct for a small 
inconsistency between the theoretical and observed 
colours. To illustrate this case, in Fig.~\ref{cmdreal_hw} we plot the corresponding
diagrams for clusters NGC~663 and NGC~884. The observed data points can be seen to follow the  curves rather closely despite appreciable extinction variations, which 
amount to  $\Delta E_{B-V}$~=0.43 and 0.30 for NGC~663 and NGC~884, respectively \citep{yadav}. These diagrams imply the distance estimates of $d$=2346 and 2402~pc
for NGC~663 and NGC~884, respectively, which agree quite well with the 2420
and 2414~pc estimate reported in the catalogue by \citet{dias02} and \citet{meynet},
respectively, although this NGC~663 distance estimate is appreciably smaller than
the  $d$=2818~pc estimate by \citet{paunzen}. 

Finally, it would be good to estimate individual $E_{r-i}$~=~$(r-i)$~-~$(r-i)_0$ 
extinction values at least for some of the stars in the cluster field. The most evident way is to determine the the $(r-i)_0$ intrinsic colour from H$_{\alpha}$~index using the appropriate isochrone (see Fig.~\ref{cciso}). To do this, we need to know the age (log~($t$)) of the cluster and
the systematic  offset ($\Delta$~H$_{\alpha}$~index) such that
H$_{\alpha}$~index (theoretical)=H$_{\alpha}$~index (observed)-$\Delta$~H$_{\alpha}$~index, both of which can be estimated by fitting the (H$_{\alpha}
$-index, $W_{ri}$) diagram (see above).   It can be seen from Fig.~\ref{cciso} that the best and most robust estimates of the $(r-i)_0$ intrinsic colour and hence the 
$E_{r-i}$~=~$(r-i)$~-~$(r-i)_0$ colour excess 
can be obtained for stars on the rising part of the ($r-i$,H$_{\alpha}$~index) colour-colour diagram with $(r-i)_0$~$\sim$~0.1--0.4 
and H$_{\alpha}$~index~$\sim$~0.0--0.16: all isochrones considered are practically indistinguishable throughout this portion of the diagram except for the two youngest 
ones (log~($t$)~=~6.0 and log~($t$)~=~6.5), which deviate from older ones at $(r-i)_0$~$\sim$~0.1--0.2 
and H$_{\alpha}$~index~$\leq$~0.05 (and even these follow closely the overall
dependence in the $(r-i)_0$~$\sim$~0.2--0.4 
and H$_{\alpha}$~index~$\sim$~0.05--0.16 domain). Furthermore, as we show in 
Section~\ref{halphaoffset}, the
offsets $\Delta$~H$_{\alpha}$~index are very similar
for all clusters and are, on the average, equal to 
$<\Delta$~H$_{\alpha}$~index$>$~=~+0.051~$\pm$~0.015, and hence
this average value can be used indiscriminately for all clusters without appreciable
loss of accuracy. Padova isochrones yield the following
common $(r-i)_0$(H$_{\alpha}$~index(observed))
calibration for stars with $(r-i)_0$~$\sim$~0.1--0.45 
and H$_{\alpha}$~index~$\sim$~0--0.17 and ages in the log~($t$)~=~7.0--8.5 interval:
\begin{equation}
(r-i)_0=+0.099+1.417x+6.59x^2-86.18x^3+424.0x^4
\label{calibr}
\end{equation}
Here  $x$ is the the estimated "theoretical" H$_{\alpha}$~index value, which in practice can be computed as 
\begin{equation}
x=H_{\alpha}~index(observed) - <\Delta~H_{\alpha}~index>,
\label{calibr1}
\end{equation}
where we can adopt $<\Delta$~H$_{\alpha}$~index$>$~=~0.051 (see Section~\ref{halphaoffset} and equation~(\ref{deltahall}) below).
Fig.~\ref{calibration} shows this calibration relation along with log~($t$)~=~6.0--8.0  isochrones.

\begin{figure}
\includegraphics[width=1.0\linewidth]{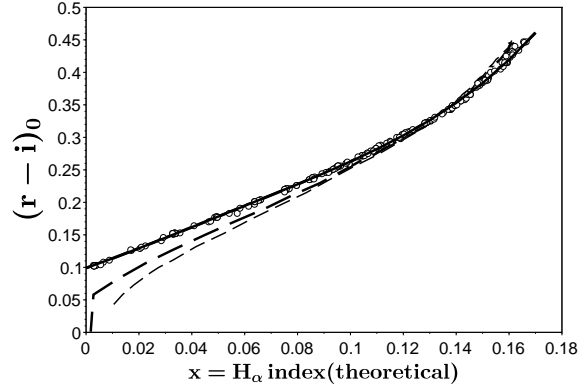}
\caption{The H$_{\alpha}$~index(theoretical)-$(r-i)_0$ diagram 
in the H$_{\alpha}$~index(theoretical)~=~0.0--0.16 and $(r-i)_0$~=~0.0--0.45
domain
with 
log~($t$)~=~7.0--8.5 isochrones (the circles) and calibration
relation~(\ref{calibr}) (the solid line). The thin and thick
dashed lines show  the  log~($t$)~=~6.0 and log~($t$)~=~6.5 isochrones.}
\label{calibration}
\end{figure}

\begin{figure}
\includegraphics[width=1.1\linewidth]{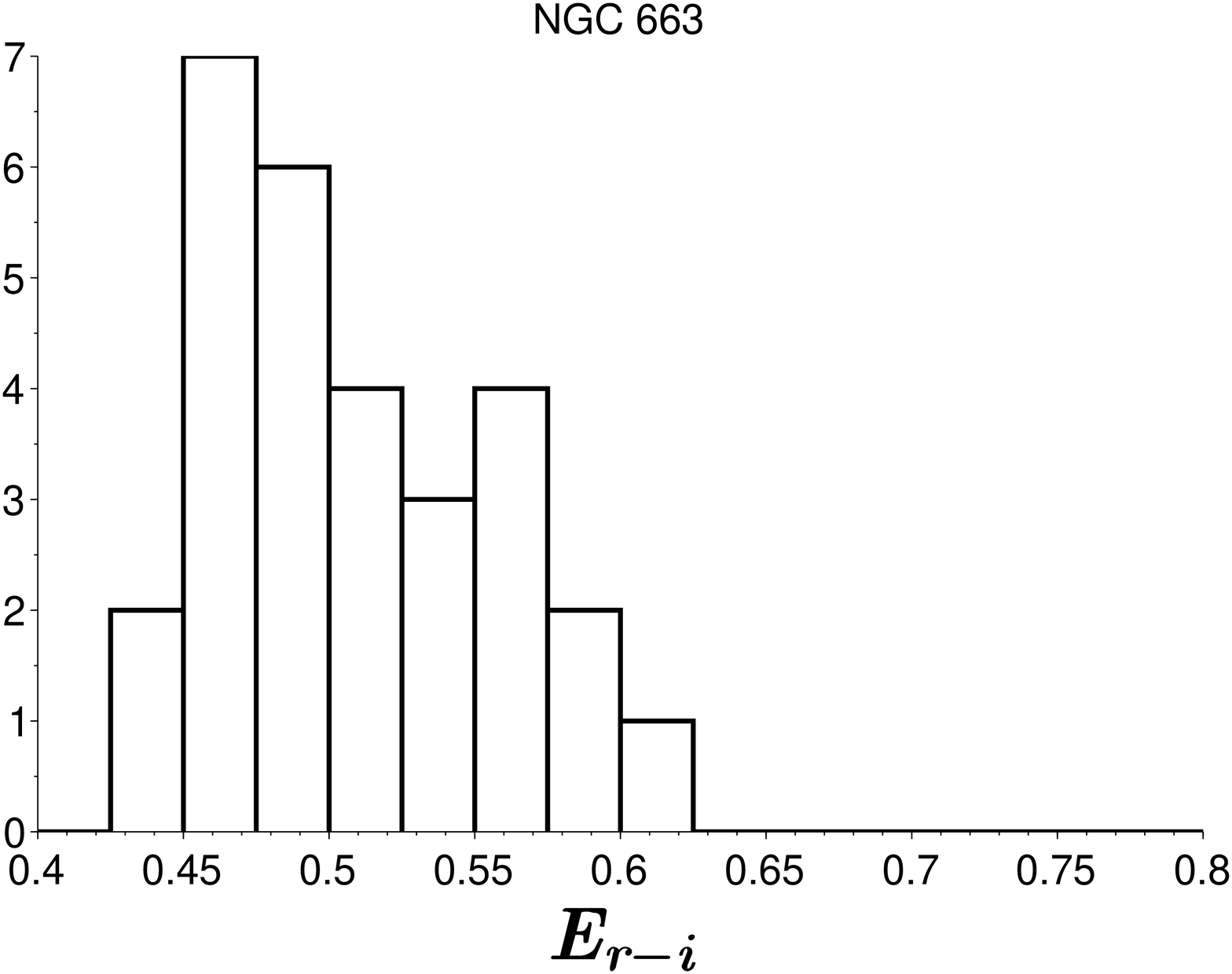}
\includegraphics[width=1.1\linewidth]{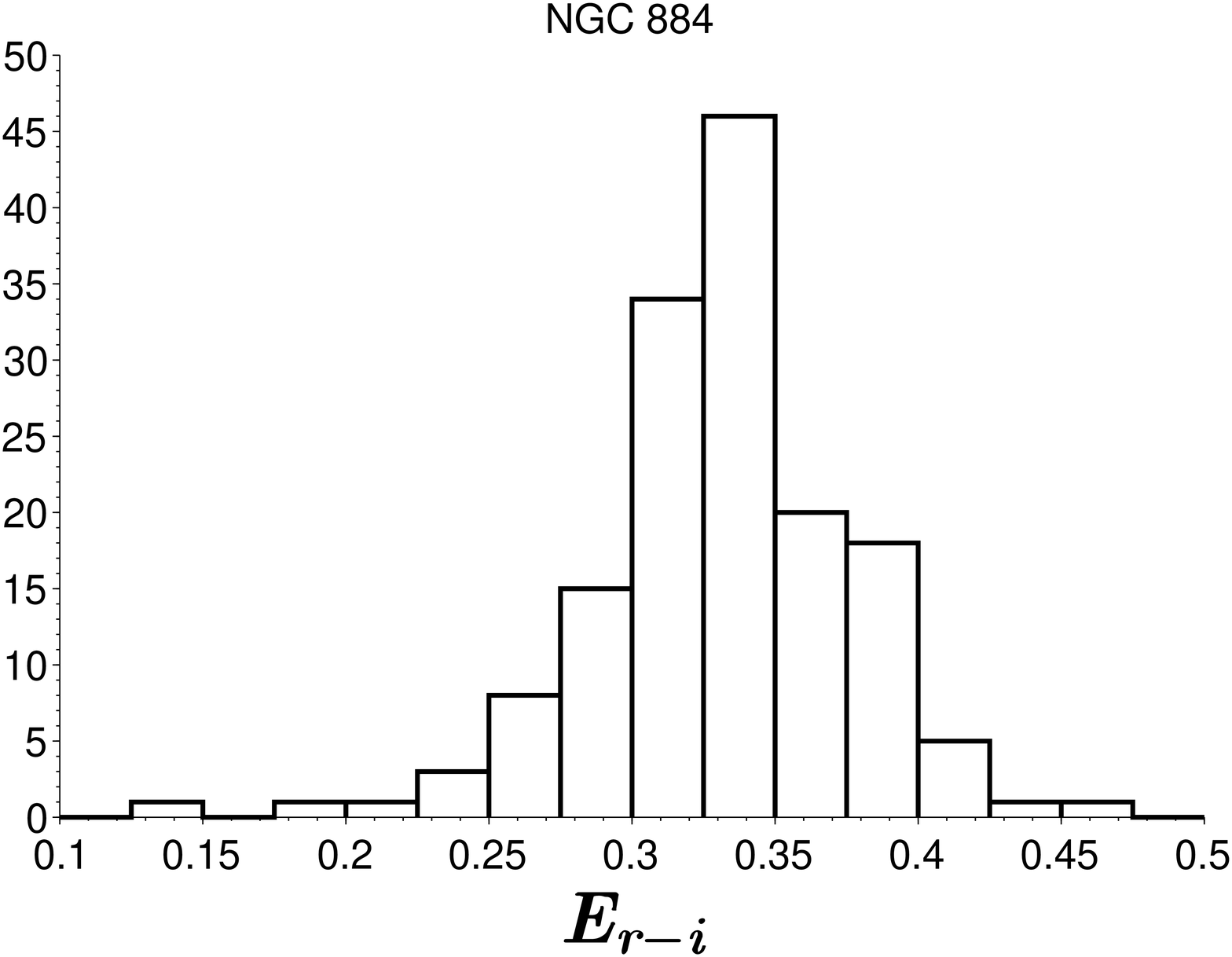}
\caption{The distribution colour excess values 
$E_{r-i}$~=~$(r-i)$~-~$(r-i)_0$ computed using calibration~(\ref{calibr}) 
for individual stars in the clusters
NGC~663 (the top panel) and NGC~884 (the bottom panel).}
\label{eri_all}
\end{figure}

To illustrate this method, we compute the color excesses $E_{r-i}$~=~$(r-i)$~-~$(r-i)_0$
of individual stars in the clusters NGC~663 and NGC~884 known for their variable
extinction (see above) and show the corresponding histograms in Fig.~\ref{eri_all}.

\subsection{Application of the maximum-likelihood technique of Naylor and Jefrries and error estimates}

Although the two diagnostic diagrams just described [($r-i$, H$_{\alpha}$~index) and 
( H$_{\alpha}$~index, $M_r$)]
allow the average color excess and distance modulus of a cluster to be determined rather accurately "by eye", such
"manual" estimates have an important drawback:
errors of the parameters derived from "eye" estimates are difficult to assess. That is why we refine 
our initial estimates by applying the formal maximum-likelihood based method proposed by \citet{naylor_jeffries}, 
\citet{jeffries}, and \citet{mayne}, which we adapt for the case of IPHAS magnitudes and colours.

The idea of this method is to use a generalized $\chi^2$ statistic proposed by the above authors,
which they call  the  $\tau^2$ statistic and which includes uncertainties in two dimensions (e.g., color and magnitude
or two different colors) and models with a two-dimensional distribution rather than just single isochronal lines. 
The $\tau^2$ value for each star measures the likelihood for the particular model based on a certain isochrone and shifted
by a certain distance modulus, extinction, and $H_{\alpha}$-index offset values to
generate a star at the observed position in the diagram taking into account the  effects of binary population and
photometric errors, and the $\tau^2$ value for the cluster is computed by summing up the corresponding
values for individual stars. The best-fit model (parametrized by reddening, age, distance and $H_{\alpha}$-index
offset or, generally, by a subset of these parameters) is supposed to be the one with minimum  $\tau^2$ and the 
corresponding parameter values are adopted as the final estimates. To compute the errors of
the inferred parameters, we use the simple bootstrap method 
with replacement \citep{efron, hastie} as described in Section~4.3 of \citet{andrae}.

\section{Investigation of ``standard" clusters}
 \label{standard_clusters}

We validated our technique of the determination of the three basic cluster 
parameters – 
colour excess and distance – by applying it to ``standard" clusters, i.e., to the clusters whose parameters 
were reliably determined in earlier studies. To this end we compare the cluster parameters – colour excess 
$E_{B-V}$, distance, and age that we determined from IPHAS  H$_{\alpha}$ri photometry with the corresponding 
estimates earlier obtained by different authors. We used the relation $E_{B-V}$ = $E_{r-i}$/0.673 to convert 
the colour excess $E_{r-i}$ 
into $E_{B-V}$, and computed the true distance modules by formula $(m-M)_0$ = $(m-M)_r$ - 3.98$E_{r-i}$. The 
coefficients 
0.673 and 3.98 are computed in accordance with the reddening law of \cite{cardelli} and adopted from 
the table of photometric system parameters provided by 
\citet{bressan} along with the isochrones. We constructed the ($r-i$,H$_{\alpha}$~index),  
(H$_{\alpha}$~index, $r$), and  (H$_{\alpha}$~index, $W_{ri}$)
diagrams for all stars inside the radii 3, 5, 
and 7~arcmin from the cluster centre depending on the angular 
size of the cluster.

We first investigated the clusters whose distance, age, and $E_{B-V}$ colour 
excess were published  by \citet{paunzen}. The above authors compiled the list of 72 open clusters 
whose parameters have been determined repeatedly by different researchers, and for which standard, 
relatively accurate (in statistical sense) values of physical parameters have been determined. 
We selected 
only those clusters among 
the 72 clusters analysed by \citet{paunzen}  that are located in the area covered by IPHAS 
and are no older than log~($t$)= 8.5 according to \citet{paunzen}. We found 
few such clusters and therefore we also added open clusters studied by \citet{meynet}, 
who also determined bona fide parameter estimates for a number of clusters and derived some calibrations for
estimating the cluster age from the blue and red MS turnoff points (here
the blue and red MS turnoff points are terms introduced by  the above authors and
graphically explained in their fig.~3). \citet{meynet} 
studied only the clusters that required no membership determination for individual stars because 
photoelectric photometry was already performed mostly for cluster members. We converted the apparent 
distance moduli published by \citet{meynet} into distances using normal extinction law $A_V$=3.1 $E_{B-V}$. 
Given that the most accurate estimates of $E_{B-V}$ colour excesses are those inferred from ($U-B$, $B-V$) 
colour-colour diagrams, we added to our sample the clusters with parameter estimates adopted from 
\citet{phelps, keller}, and \citet{lata}. The above authors first determined the $E_{B-V}$ colour 
excess by fitting cluster MS on the ($U-B$, $B-V$) diagram to the standard ZAMS of 
\citet{schmidtkaler} or to the youngest isochrones of 
\citet{girardi}, and then determined the distance modulus and age from the colour-magnitude 
diagrams.

We  proceed as follows.
\begin{itemize} 
\item For each cluster we first estimate its average
colour excess $E_{r-i}$ and $\Delta$~H$_{\alpha}$~index offset by fitting the log~($t$)~=7.2 isochrone (with the age approximatively
midway between our age interval boundaries (log~($t$)~=~6.0 and log~($t$)~=~8.5) to the
($r-i$,H$_{\alpha}$~index) diagram based on IPHAS data (see Fig.~\ref{cc7790}). 

\item We then use this  initial 
$E_{r-i}$ estimate to fit the appropriate isochrone to  the ($r-i$)--$r$ diagram based on 
IPHAS and transformed APASS data (the latter are used for stars that are too bright  
and hence saturated in the IPHAS frames) and infer a preliminary age estimate  (log~(t$_0$) -- see Fig.~\ref{cmd7790_rir}).

\item We then return to the first step and refine the colour excess $E_{r-i}$ and 
$\Delta$~H$_{\alpha}$~index offset by fitting the ($r-i$,H$_{\alpha}$~index) diagram to  the isochrone with the preliminary 
age estimate (instead of log~($t$)=7.2).

\item We now use the refined $E_{r-i}$  estimate to fit the 
appropriate isochrone to the ($r-i$)--$r$ diagram and determine the final age (log~(t$_1$)) and apparent distance modulus ($DM_r$) estimates and compute the first
distance (D$_1$) estimates by dereddening the  inferred apparent distance modulus.

\item We then fit the (H$_{\alpha}$~index, $r$) diagram for cluster stars 
with the (H$_{\alpha}$~index, $M_r$) isochrone with fixed log~(t$_1$) 
(see Fig.~\ref{cmd7790_hr}) and determine another distance (D$_2$) and another $\Delta$~H$_{\alpha}$~index offset estimate.

\item We finally fit an appropriate isochrone to the reddening-free 
(H$_{\alpha}$~index, $W_{ri}$) diagram (see Fig.~\ref{cmdreal_hw}) 
with the true distance modulus, age, and
$\Delta$~H$_{\alpha}$~index offset treated as free parameters to determine
the third distance (D$_3$) and the  third $\Delta$~H$_{\alpha}$~index offset estimates and the
second age estimate (log~(t$_2$)) for the cluster.

\item Finally, we determine the monochromatic extinction $A_0$ at 
5495\AA~from the 3D map of \citet{sale} and the $A_V-A_0$ difference,
where $A_V$~=~3.1$E_{B-V}$ (we compute $E_{B-V}$~=~$E_{r-i}$/0.673) 
is our total V-band extinction estimate.

\end{itemize}

\section{Results}
\label{results}

The results for all 18 ``standard" clusters are summarized in Table~2.
Here column~1 gives the cluster name; column~2, our final $E_{B-V}$
colour excess estimate  ($E_{B-V}$~=~$E_{r-i}$/0.673) (at the top of the cell) and published 
$E_{B-V}$ estimates (at the bottom of the cell); column~3 gives our three distance estimates based on 
three photometric diagrams (indicated in the  parentheses) at the top 
of the cell followed by published
distance estimates at the bottom of the cell;  column~4 lists our two age estimates based on two photometric 
diagrams (indicated in the parentheses) at the top of the cell followed by published age estimates at the bottom of the cell; column~5 gives the corresponding references to the published colour excess, distance, and age values listed in columns 2--4; column~6 gives
the monochromatic extinction $A_0$ at 
5495\AA~from the 3D map of \citet{sale},  and column~7, the $A_V-A_0$ difference.

For comparison, we also list in Table~2 the cluster parameters from the 
recent version of DAML02 open cluster catalog by
\citet{dias02} (columns~2--4, the bottom part of the cells) except for the clusters for which DAML02 adopted the color excess, 
distance, and age from the studies cited above. 

\begin{figure}
\includegraphics[width=1.1\linewidth]{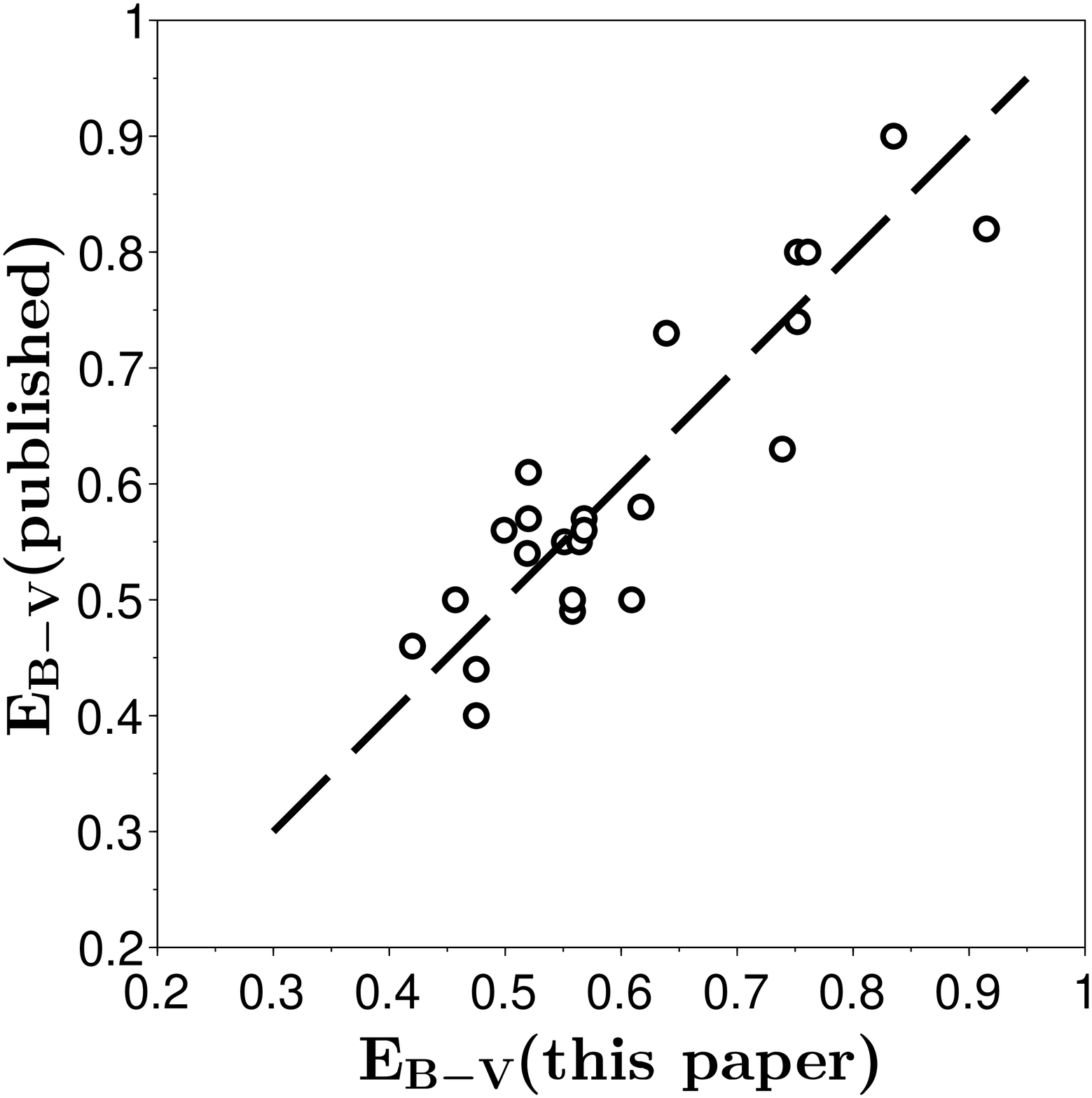}
\includegraphics[width=1.1\linewidth]{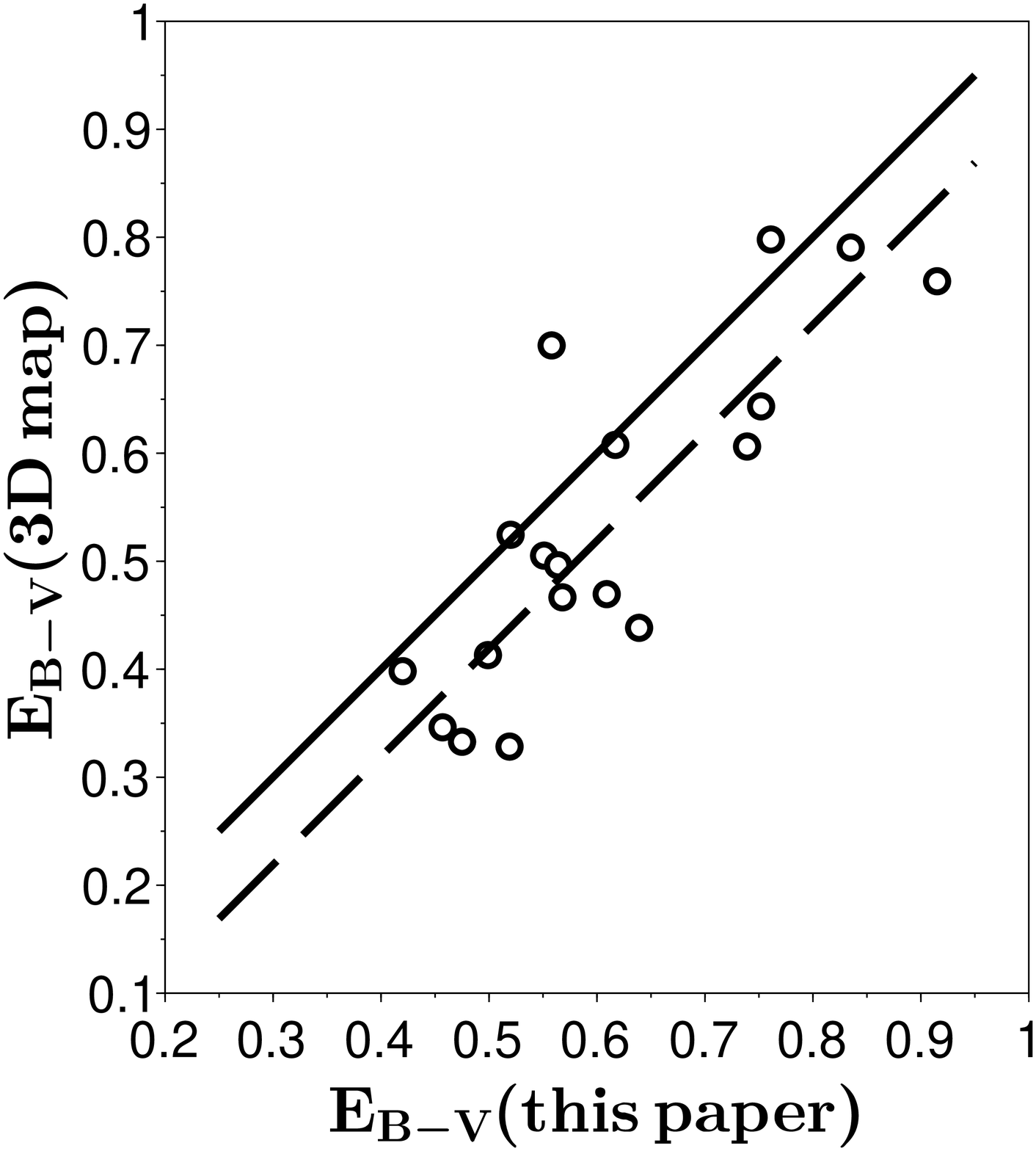}
\caption{Top panel: comparison of our $E_{B-V}$ colour excess estimates for ``standard"
clusters with published values. The dashed line corresponds to
$E_{B-V}$(published)~=~$E_{B-V}$(this paper). Bottom panel: comparison of our $E_{B-V}$ colour excess estimates for ``standard"
clusters with those implied by the 3D extinction map by  \citet{sale}. The dashed line corresponds to $E_{B-V}$(3D map)~=~$E_{B-V}$(this paper)-0.081 relation.}
\label{ebvcompar}
\end{figure}

Our extinction and distance estimates agree well with those reported in most 
of the “standard” cluster studies - \citet{paunzen, meynet, phelps, lata, keller}:

$$
<E_{B-V}(Publ)-E_{B-V}> = -0.005~\pm~0.012, 
$$

$$
\sigma~(E_{B-V}(Publ)-E_{B-V})=0.061,
$$
where $E_{B-V}$ and $E_{B-V}(Publ)$ 
are our inferred extinction values and the original 
published estimates of the above authors, respectively (see Table~\ref{compar}
and Fig.~\ref{ebvcompar}).

On the other hand, our extinction values deviate appreciably
from those implied by the 3D map of \citet{sale}: we have in the 
case of the $A_V-A_0$ difference, 
where $A_V$ is our total 
V-band extinction estimate: 
$$
<A_V-A_0> = +0.251 ~\pm~ 0.018, \sigma~(A_V-A_0) = 0.077, 
$$
or, in terms of colour excesses 
$$
<E_{B-V}-(A_0/3.1)> = +0.081 ~\pm~ 0.006, 
$$

$$
\sigma~(E_{B-V}-(A_0/3.1)) = 0.025
$$
(see Table~\ref{compar}
and Fig.~\ref{ebvcompar}). 
This difference may be due to the limits imposed by the finite sampling 
of the extinction map of 
\citet{sale} mentioned above: here we adopted the $A_0$ values corresponding to sky 
locations 5--7~arcmin from the cluster centre in most of the cases. The systematically 
higher values of our extinction estimates may be due to internal extinction within the 
cluster. 

\begin{figure}
\includegraphics[width=1.1\linewidth]{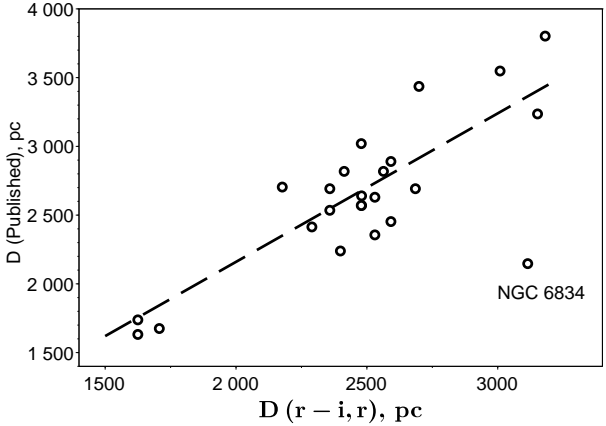}
\caption{Comparison of our distance estimates $D(r-i,r)$ with published values. The dashed line is the $D(Published)$~=~1.080$D(r-i,r)$ relation inferred from all
clusters except NGC~6834.}
\label{discomp}
\end{figure}

Our distance estimates 
based on the $(r-i, r)$, (H$_{\alpha}$~index, $r$),
and (H$_{\alpha}$~index, $W_{ri}$) diagrams agree well with each other:
$$
<D(r, H_{\alpha}~index)/D(r-i,r)> = 0.980~\pm~0.009, 
$$

$$
\sigma~(D(r, H_{\alpha}~index)/D(r-i,r))=0.037,
$$

and
$$
<(D(W_{ri}, H_{\alpha}~index)/D(r-i,r)>~=~1.000~\pm~0.008 
$$

$$
\sigma~((D(W_{ri}, H_{\alpha}~index)/D(r-i,r))=0.035.
$$
They also agree well with earlier published values except for the 
NGC6834 cluster for which we obtain a significantly greater distance (3115~$\pm$~63,
3115~$\pm$~120, and 2950~$\pm$~59 based on $(r-i, r)$, (H$_{\alpha}$~index, $r$),
and (H$_{\alpha}$~index, $W_{ri}$) diagrams, respectively) than inferred 
by \citet{paunzen} (2147~$\pm$~59) (see Fig.~\ref{discomp}):
$$
<D(Published)/D(r-i,r)> = 1.080~\pm~0.020, 
$$

$$
\sigma~(D(Published)/D(r-i,r))=0.098,
$$

$$
<D(Published)/D(r, H_{\alpha}~index)> = 1.105~\pm~0.026, 
$$

$$
\sigma~(D(Published)/D(r, H_{\alpha}~index))=0.123,
$$
and
$$
<D(Published)/D(W_{ri}, H_{\alpha}~index)> = 1.075~\pm~0.020, 
$$

$$
\sigma~(D(Published)/D(W_{ri}, H_{\alpha}~index))=0.094,
$$

where $D(Published)$ are published distance estimates adopted
from \citet{paunzen, meynet, phelps, lata, keller} and $D(r-i,i)$,
$D(r, H_{\alpha}~index)$, and $D(W_{ri}, H_{\alpha}~index)$ are our 
distance estimates based on the $(r-i,i)$,
$(r, H_{\alpha}~index)$, and $D(W_{ri}, H_{\alpha}~index)$ diagrams, respectively
(see Table~\ref{compar}).

The overall good agreement of our inferred colour excesses and distances 
with earlier estimates suggests that the technique presented in this work can be used to 
estimate the two above parameters for those clusters that do not have observations in the $U$ band. 

\begin{figure}
\includegraphics[width=1.1\linewidth]{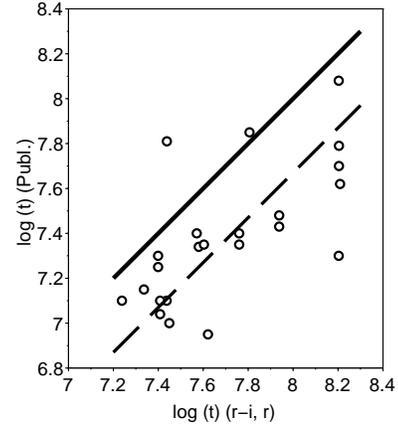}
\caption{Comparison of our age estimates log~($t(r-i,r)$) with published values. 
The solid and dashed lines show the identity (log~(t(Published))~=~log~(t$(r-i,r)$))
and log~(t(Published))~=~log~(t $(r-i,r)$)--0.33 relations, respectively.}
\label{agecomp}
\end{figure}

However, our age estimates based on the $(r-i,i)$ and $(W_{ri}, H_{\alpha}~index)$
diagrams do not agree so well with 
published values:
$$
<log~(t)(Published)-log~(t) (r-i,r)> = +0.33~\pm~0.06, 
$$

$$
\sigma~(log~(t)(Published)-log~(t) (r-i,r))=0.27,
$$
and
$$
<log~(t)(Published)-log~(t) (W_{ri}, H_{\alpha}~index)> = +0.46~\pm~0.07, 
$$

$$
\sigma~(log~(t)(Published)-log~(t) (W_{ri}, H_{\alpha}~index))=0.36,
$$
which is to be expected given the saturation of bright stars in IPHAS images.
As expected, the results based on the $(r-i,r)$ diagram are slightly better than 
those based on the $(W_{ri}, H_{\alpha}~index)$ diagram because the former includes 
the data adopted from APASS for relatively bright evolved stars 
(see Table~\ref{comparage} and Fig.~\ref{agecomp}).

\begin{table}
  \centering
%  \begin{minipage}{140mm}
  \caption{Intrinsic $r-i$  colour of the H$_{\alpha}$~index minimum feature as a function of age and metallicity.}\label{agemet}
\begin{tabular}{r r r r}
\hline
     & \multicolumn{3}{c}{r-i  at minimum H$_{\alpha}$~index} \\
 Age, log~($t$) & Z = 0.0048     & Z = 0.0152    &  Z = 0.0481    \\
            & [Fe/H] = -0.5 & [Fe/H] = 0.0 & [Fe/H] = +0.5 \\
\hline
6.0 & 0.032 & 0.027 & 0.032 \\
6.5 & 0.024 & 0.024 & 0.027 \\
7.0 & 0.020 & 0.020 & 0.021 \\
7.5 & 0.022 & 0.024 & 0.026 \\
8.0 & 0.022 & 0.024 & 0.027 \\
8.5 & 0.026 & 0.031 & 0.032 \\
\hline
 Mean &       0.028 &       0.025 &        0.026    \\
      & $\pm$~0.004 & $\pm$~0.004 &  $\pm$~0.004    \\
\hline
 Overall mean & \multicolumn{3}{c}{0.026~$\pm$~0.004}  \\
\hline
\end{tabular}
%\end{minipage}
\end{table}

\begin{center}
\onecolumn
\begin{longtable}{|l|r|l|l|l|r|l|}
\caption[Comparison of parameter estimates for ``standard" open clusters.]{Comparison of parameter estimates for ``standard" open clusters.}\label{standard} \\
\hline
Cluster & $E_{B-V}$ & d$^*$, pc & log($t$)$^*$ & reference$^{**}$ & A$_{0}$ & A$_{v}$-A
$_{0}$ (${^{***}}$) \\
\hline
\endfirsthead
\hline
Cluster & $E_{B-V}$ & d$^*$, pc & log($t$)$^*$ & reference$^{**}$ & A$_{0}$ & A$_{v}$-A
$_{0}$ (${^{***}}$) \\
\hline
\endhead
  \multicolumn{7}{l}{{$^*$ ~The colour-magnitude diagram used to compute $d$ and log~($t$) is indicated in the parentheses. } } \\
\multicolumn{7}{l}{{{${^{**}}$ PN06: \citet{paunzen}; L14: \citet{lata}; DAML02: \citet{dias02};  }}} \\
\multicolumn{7}{l} {{{M93: \citet{meynet}; PJ94: \citet{phelps}; K05: \citet{keller}.}}} \\     
\multicolumn{7}{l}{{${^{***}}$ A$_{v}$~=~3.1$E_{B-V}$.}} \\   
  \multicolumn{7}{l}{{Continued on Next Page\ldots}} \\  
\endfoot 
%This is the footer for the last page of the table...
\multicolumn{7}{l}{{$^*$ ~The colour-magnitude diagram used to compute $d$ and log~($t$) is indicated in the parentheses. } } \\
\multicolumn{7}{l}{{{${^{**}}$ PN06: \citet{paunzen}; L14: \citet{lata}; DAML02: \citet{dias02}; M93: \citet{meynet}; }}}\\
\multicolumn{7}{l} {{{PJ94: \citet{phelps}; K05: \citet{keller}.}}} \\     
%%\multicolumn{7}{l}{{${^{**}}$ PN06: \citet{paunzen}; DAML02: \citet{dias02}; M93: \citet{meynet}; PJ94: \citet{phelps}; K05: \citet{keller}}\\   
\multicolumn{7}{l}{{${^{***}}$ A$_{v}$~=~3.1$E_{B-V}$.}} \\ 
\endlastfoot

Be 7 & 0.752 & 2480~$\pm$~87 ($r-i,r$) &  7.44~$\pm$~0.16 ($r-i,r$)   & This          & 1.99 & +0.34 \\
     &  $\pm$~0.009  & 2513~$\pm$~42 (H$_{\alpha}$~index, $r$) &         & paper &$\pm$~0.04  &  \\
     &                   & 2484~$\pm$~140 (H$_{\alpha}$~index, $W$) &  7.41~$\pm$~0.24 (H$_{\alpha}$~index, $W$)       &               &  &  \\  
     &                   &  &        &               &  &  \\            
     & 0.80 & 2570 & 6.60    & PJ94 &  \\
     & 0.74 & 2640 & 7.1     & L14   &                  &  \\

\hline
Be 62 & 0.915 & 2177~$\pm$~34 ($r-i,r$) & 7.45~$\pm$~0.10 ($r-i,r$)    & This  & 2.35  & +0.49 \\
     &  $\pm$~0.015  & 2513~$\pm$~42 (H$_{\alpha}$~index, $r$) &         & paper &$\pm$~0.04  &  \\
     &                   & 2250~$\pm$~45 (H$_{\alpha}$~index, $W$) &  7.91~$\pm$~0.18 (H$_{\alpha}$~index, $W$)      &               &  &  \\     
     &                   &  &        &               &  &  \\                 
      & 0.82 & 2704 & 7.00 & PJ94 & &  \\
      & 0.85 & 2320 & 7.2  & DAML02  & &  \\
\hline
King 12 & 0.618 & 2592~$\pm$~46 ($r-i,r$) & 7.41~$\pm$~0.03 ($r-i,r$)     & This & 1.88 & +0.04 \\
     &    $\pm$~0.007  & 2339~$\pm$~64 (H$_{\alpha}$~index, $r$) &         & paper & $\pm$~0.06 &  \\
     &                   & 2595~$\pm$~87 (H$_{\alpha}$~index, $W$) &  7.22~$\pm$~0.04 (H$_{\alpha}$~index, $W$)      &               &  &  \\     
     &                   &  &        &               &  &  \\                 
        & 0.58 & 2453~$\pm$~120 & 7.04 & PN06 &                        &     \\
        & $\pm$~0.03 &                &      &      &                        &     \\        
        & 0.58            & 2890           & 7.1  & L14  &                        &     \\
\hline
NGC 103 & 0.551 & 3181~$\pm$~38 ($r-i,r$) &  8.20~$\pm$~0.06  ($r-i,r$)   & This & 1.57  & +0.14 \\
     &    $\pm$~0.010 & 3307~$\pm$~110 (H$_{\alpha}$~index, $r$) &         & paper & $\pm$~0.04  &  \\
     &                   & 3199~$\pm$~73 (H$_{\alpha}$~index, $W$) &  8.26~$\pm$~0.10 (H$_{\alpha}$~index, $W$)      &               &  &  \\     
     &                   &  &        &               &  &  \\                 
        & 0.55  & 3802 & 7.30  & PJ94 &  &  \\
        & 0.406 & 3026 & 8.126 & DAML02  &  &  \\
\hline
NGC 129 & 0.569  & 1625~$\pm$~14 ($r-i,r$) & 8.20~$\pm$~0.04 ($r-i,r$)  & This & 1.45 & +0.31 \\
     &  $\pm$~0.018 & 1605~$\pm$~24 (H$_{\alpha}$~index, $r$) &         & paper & $\pm$~0.04 &  \\
     &                   & 1690~$\pm$~20 (H$_{\alpha}$~index, $W$) &  8.30~$\pm$~0.01 (H$_{\alpha}$~index, $W$)      &               &  &  \\     
     &                   &  &        &               &  &  \\                 
        & 0.56 & 1632~$\pm$~56 & 7.79   & PN06 & &  \\
        & $\pm$~0.03 &                &      &      &                        &     \\                
        & 0.57            & 1738          & 7.70   & PJ94 & &  \\
        & 0.548           & 1625          & 7.886  & DAML02  & &  \\
\hline
%%\pagebreak
NGC 436 & 0.458 & 3152~$\pm$~34 ($r-i,r$) &  8.21~$\pm$~0.04 ($r-i,r$)     & This & 1.07 & +0.35 \\
     &   $\pm$~0.006  & 3089~$\pm$~60 (H$_{\alpha}$~index, $r$) &         & paper & $\pm$~0.04  &  \\
     &                   & 3070~$\pm$~26 (H$_{\alpha}$~index, $W$) &  8.10~$\pm$~0.08 (H$_{\alpha}$~index, $W$)      &               &  &  \\     
     &                   &  &        &               &  &  \\                 
        & 0.50  &  3236 & 7.62  & PN06 &  &  \\
        & 0.460 &  3014 & 7.926 & DAML02  &  &  \\
\hline
NGC 457 & 0.559  & 2479~$\pm$~39 ($r-i,r$) & 7.40~$\pm$~0.07 ($r-i,r$)     & This & 1.43 & +0.28 \\
     &   $\pm$~0.006                & 2331~$\pm$~26 (H$_{\alpha}$~index, $r$) &         & paper & $\pm$~0.01  &  \\
     &                   & 2476~$\pm$~30 (H$_{\alpha}$~index, $W$) &  8.10~$\pm$~0.09 (H$_{\alpha}$~index, $W$)      &               &  &  \\     
     &                   &  &        &               &  &  \\                 
        & 0.50  & 2570 & 7.25  & M93  &  &  \\
        & 0.49  & 3020 & 7.30  & PJ94 &  &  \\
        & 0.472 & 2429 & 7.324 & DAML02  &  &  \\
\hline
NGC 581 & 0.475 & 2358~$\pm$~20 ($r-i,r$) &  7.76~$\pm$~0.06 ($r-i,r$) & This & 1.03 & +0.44 \\
     &    $\pm$~0.006 & 2207~$\pm$~150 (H$_{\alpha}$~index, $r$) &        & paper & $\pm$~0.03  &  \\
     &                   & 2427~$\pm$~150 (H$_{\alpha}$~index, $W$) &  8.22~$\pm$~0.14 (H$_{\alpha}$~index, $W$)     &               &  &  \\     
     &                   &  &        &               &  &  \\                 
        & 0.40   & 2535  & 7.40  & M93  &  &  \\
        & 0.44   & 2692  & 7.35  & PJ96 &  &  \\
        & 0.382  & 2194  & 7.336 & DAML02  &  &  \\
\hline
NGC 654 & 0.835 & 2685~$\pm$~75 ($r-i,r$) & 7.57~$\pm$~0.14 ($r-i,r$) & This & 2.45 & +0.14 \\
     &   $\pm$~0.013  & 2507~$\pm$~54 (H$_{\alpha}$~index, $r$) &         & paper & $\pm$~0.05 &  \\
     &                   & 2704~$\pm$~73 (H$_{\alpha}$~index, $W$) &  8.10~$\pm$~0.13 (H$_{\alpha}$~index, $W$)       &               &  &  \\     
     &                   &  &        &               &  &  \\                 
        & 0.90 & 2692  & 7.40  & PJ96 &  &  \\
        & 0.82 & 2410  & 7.0   & DAML02  &  &  \\
\hline
NGC 659 & 0.738 & 2699~$\pm$~94 ($r-i,r$) & 7.58~$\pm$~0.06 ($r-i,r$) & This & 1.88 & +0.41 \\
     &    $\pm$~0.018 & 2699~$\pm$~94 (H$_{\alpha}$~index, $r$) &        & paper & $\pm$~0.03 &  \\
     &                   & 2767~$\pm$~42 (H$_{\alpha}$~index, $W$) &  7.43~$\pm$~0.28 (H$_{\alpha}$~index, $W$)       &               &  &  \\     
     &                   &  &        &               &  &  \\                 
        & 0.63   & 3436 & 7.34  & PJ96 &  &  \\
        & 0.652  & 1938 & 7.548 & DAML02 &  &  \\
\hline
NGC 663 & 0.761 & 2413~$\pm$~73 ($r-i,r$) & 7.60~$\pm$~0.08 ($r-i,r$) & This & 2.48 & -0.12 \\
     &    $\pm$~0.015 & 2408~$\pm$~71 (H$_{\alpha}$~index, $r$) &         & paper & $\pm$~0.09 &  \\
     &                   & 2402~$\pm$~31 (H$_{\alpha}$~index, $W$) &  7.77~$\pm$~0.12 (H$_{\alpha}$~index, $W$)      &               &  &  \\     
     &                   &  &        &               &  &  \\                 
        & 0.80 & 2818 & 7.35    & PJ06 &  & \\
        & 0.80 & 2420 & 7.4     & DAML02  &  &  \\
\hline
NGC 869 & 0.519 & 2399~$\pm$~66 ($r-i,r$) & 7.24~$\pm$~0.04 ($r-i,r$) & This & 1.02 & +0.59 \\
     &     $\pm$~0.016 & 2297~$\pm$~26 (H$_{\alpha}$~index, $r$) &         & paper & $\pm$~0.02 &  \\
     &                   & 2391~$\pm$~60 (H$_{\alpha}$~index, $W$) &  7.28~$\pm$~0.06 (H$_{\alpha}$~index, $W$)      &               &  &  \\     
     &                   &  &        &               &  &  \\                 
        & 0.54   & 2239 & 7.1   & K05 &  &  \\
        & 0.575  & 2079 & 7.069 & DAML02 &  &  \\
\hline
NGC 884 & 0.499  & 2290~$\pm$~28 ($r-i,r$) & 7.34~$\pm$~0.12 ($r-i,r$) & This & 1.28 & +0.27 \\
     &    $\pm$~0.009 & 2188~$\pm$~29 (H$_{\alpha}$~index, $r$) &         & paper & $\pm$~0.01 &  \\
     &                   & 2346~$\pm$~25 (H$_{\alpha}$~index, $W$) &  7.31~$\pm$~0.09 (H$_{\alpha}$~index, $W$)      &               &  &  \\     
     &                   &  &        &               &  &  \\                 
        & 0.56  & 2414  & 7.15  & M93 &  &  \\
        & 0.56  & 2940  & 7.1   & DAML02 &  &  \\
\hline
NGC 6834 & 0.639 & 3115~$\pm$~63 ($r-i,r$)  & 7.44~$\pm$~0.12 ($r-i,r$) & This & 1.36 & +0.62 \\
     &     $\pm$~0.012 & 3115~$\pm$~120 (H$_{\alpha}$~index, $r$) &         & paper & $\pm$~0.05 &  \\
     &                   & 2950~$\pm$~36 (H$_{\alpha}$~index, $W$) &  7.91~$\pm$~0.10 (H$_{\alpha}$~index, $W$)       &               &  &  \\     
         & 0.73 & 2147~$\pm$~59 & 7.81  & PN06 &  &  \\
     &     $\pm$~0.03           &  &        &               &  &  \\                     
\hline
%%\pagebreak
NGC 6871 & 0.421 & 1707~$\pm$~43 ($r-i,r$) & 7.62~$\pm$~0.25 ($r-i,r$) & This & 1.24 & +0.07 \\
     &     $\pm$~0.007 & 1617~$\pm$~73 (H$_{\alpha}$~index, $r$) &         & paper & $\pm$~0.09 &  \\
     &                   & 1615~$\pm$~41 (H$_{\alpha}$~index, $W$) &  8.02~$\pm$~0.04 (H$_{\alpha}$~index, $W$)    &               &  &  \\     
     &                   &  &        &               &  &  \\                 
        & 0.46 & 1675~$\pm$~130 & 6.95 & PN06 &  &  \\
        & $\pm$~0.03 &                &      &      &                        &     \\                 
\hline
NGC 7790 & 0.565 & 3006~$\pm$~90 ($r-i,r$) & 7.81~$\pm$~0.23 ($r-i,r$) & This & 1.54 & +0.21 \\
     &     $\pm$~0.006 & 2793~$\pm$~72 (H$_{\alpha}$~index, $r$) &       & paper & $\pm$~ 0.04 &  \\
     &                   & 2858~$\pm$~44 (H$_{\alpha}$~index, $W$) &  7.86~$\pm$~0.17 (H$_{\alpha}$~index, $W$)       &               &  &  \\     
     &                   &  &        &               &  &  \\                 
         & 0.55   & 3548 & 7.85  & PJ94 &  &  \\
         & 0.531  & 2944 & 7.749 & DAML02 &  &  \\
\hline
Stock 24 & 0.609 & 2563~$\pm$~28 ($r-i,r$) & 8.20~$\pm$~0.02 ($r-i,r$) & This  & 1.46  & +0.43 \\
     &     $\pm$~0.015 & 2570~$\pm$~50 (H$_{\alpha}$~index, $r$) &   & paper & $\pm$~0.06 &  \\
     &                   & 2750~$\pm$~59 (H$_{\alpha}$~index, $W$) &  8.27~$\pm$~0.03 (H$_{\alpha}$~index, $W$)     &               &  &  \\     
     &                   &  &        &               &  &  \\                 
         & 0.50 & 2818 & 8.08  & PJ94 &  &  \\
\hline
Trumpler 1 & 0.520 & 2530~$\pm$~42 ($r-i,r$) & 7.94~$\pm$~0.18 ($r-i,r$) & This  & 1.63  & -0.02 \\
     &       $\pm$~0.082 & 2903~$\pm$~160 (H$_{\alpha}$~index, $r$) &         & paper & $\pm$~0.03  &  \\
     &                   & 2437~$\pm$~37 (H$_{\alpha}$~index, $W$) &  7.43~$\pm$~0.10 (H$_{\alpha}$~index, $W$)      &               &  &  \\     
     &                   &  &        &               &  &  \\                 
        & 0.57 & 2356~$\pm$~510 & 7.48 & PN06 & &  \\
        & $\pm$~0.04 &                &      &      &                        &     \\                   
           & 0.61            & 2630           & 7.43 & PJ94  & &  \\
\hline           
\label{standard}
\end{longtable}
\end{center}
\twocolumn

\section{The systematic H$_{\alpha}$~index offset}
\label{halphaoffset}

As we already pointed out above, the observed H$_{\alpha}$~index values differ
systematically from those implied by the corresponding isochrones.
Table~\ref{deltah} lists the $\Delta$~H~$_{\alpha}$-index estimates for
18 ``standard" clusters based on ($r-i$,H$_{\alpha}$~index), (H$_{\alpha}$~index, $r$),
and (H$_{\alpha}$~index, $W$) diagrams (columns~2 to 4) and the weighted averages
of these estimates for each cluster (column~5). All three diagrams yield highly consistent $\Delta$~H~$_{\alpha}$-index values
for each cluster with the average difference between the values inferred from
the ($r-i, H_{\alpha}~index$) and ($H_{\alpha}~index, r$) diagrams equal to +0.004~$\pm$~0.002
with a scatter of 0.009 and the average difference between the values inferred from
the ($H_{\alpha}~index, r$) and ($H_{\alpha}~index, W$) diagrams equal to +0.001~$\pm$~0.001
with a scatter of 0.005.

\begin{figure}
\includegraphics[width=1.1\linewidth]{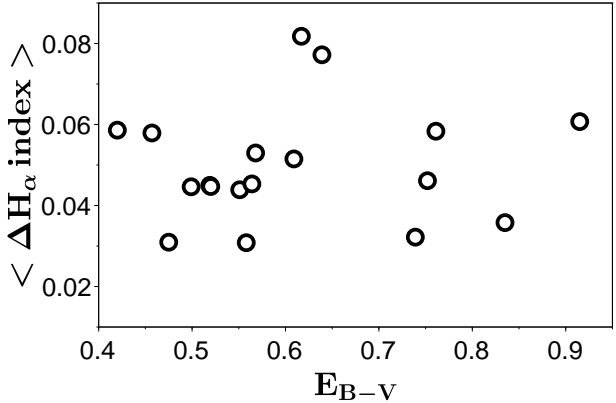}
\caption{Plot of $\Delta$~H~$_{\alpha}$-index offset computed as the average 
of the estimates based on three photometric diagrams versus the
$E_{B-V}$ colour excess. The offsets show no trend with colour excess:
$<\Delta$~H~$_{\alpha}$-index$>$~=~+0.049($\pm$~0.005)+0.007($\pm$~0.026)($E_{B-V}$-0.5).}
\label{deltah_trend}
\end{figure}

Furthermore, the scatter of $\Delta$~H~$_{\alpha}$-index values for different clusters
is also rather small:
$$
<\Delta~H~_{\alpha}~index)> = +0.051
$$
\begin{equation}
\sigma (\Delta~H~_{\alpha}~index)=0.014.
\label{deltahall}
\end{equation}
One possible cause of such a systematic difference may be a mismatch between the 
zero points of the $r,h,H_{\alpha}$ magnitudes used in isochrone computations and the 
corresponding zero points of actually observed magnitudes. Another explanation is
that the adopted reddening law is not correct in the wavelength interval between the
$r$ and $i$ bands and hence the computed H$_{\alpha}$~index is not extinction independent.
However, in this latter case there should be a trend of $\Delta$~H~$_{\alpha}$-index with
extinction $E_{B-V}$, which is actually absent (see Fig.~\ref{deltah_trend}). 
Hence the first explanation seems more plausible and therefore the same $\Delta$~H~$_{\alpha}$-index value 
($<\Delta~H_{\alpha}~index>$~=~+0.051) can be used in calibration~(\ref{calibr})--(\ref{calibr1})
in Section~\ref{nonuniform} above.

\begin{table*}
\begin{flushleft}
  \caption{Comparison of extinction and distance estimates determined in this paper 
for ``standard" clusters with earlier published values .}\label{compar}
\begin{tabular}{|l|r|r|r|r|r|r|r|r|}
\hline
Reference & $E_{B-V}$-          & $<$d/d(Publ)$>$ & $\sigma$~(d/d(Publ)) & $<$d/d(Publ)$>$ & $\sigma$~(d/d(Publ)) & $<$d/d(Publ)$>$ & $\sigma$~(d/d(Publ)) &  Number \\
          & $E_{B-V}$(Publ)     &   \multicolumn{2}{|c|}{($r-i$,  r }  & \multicolumn{2}{|c|}{(H$_{\alpha}$~index, $r$)} & \multicolumn{2}{|c|}{(H$_{\alpha}$~index, $W$)} & of clusters\\
\hline
PN06     &  -0.027      &  1.12         & 0.19 & 1.11 & 0.22 & 1.09 & 0.16 & 5 \\
         &  $\pm$~0.022 &  $\pm$~ 0.08  &      & $\pm$~ 0.10 & & $\pm$~ 0.07 & &  \\
         &              &               &      &      &      &      &      & \\  
PN06     &  -0.011      &  1.04         & 0.04 & 1.03 & 0.13 & 1.02 & 0.04 & 4 \\
without  &  $\pm$~0.021 &  $\pm$~ 0.02  &      & $\pm$~ 0.07 & & $\pm$~ 0.02 &  & \\
NGC~6834 &              &               &      &      &      &      &      & \\                 
         &              &               &      &      &      &      &      & \\
PJ94     &  +0.011      &  0.89         & 0.07 & 0.89 & 0.07 & 0.90 & 0.07 & 13 \\
         &  $\pm$~0.019 &  $\pm$~ 0.02  &      & $\pm$~ 0.02 & & $\pm$~ 0.02 & &       \\
         &              &               &      &      &      &      &      & \\        
M93      &  +0.024      &  0.95         & 0.03 & 0.89 & 0.02 & 0.96 & 0.01 & 3 \\
         &  $\pm$~0.042 &  $\pm$~ 0.01  &      & $\pm$~ 0.01 & & $\pm$~ 0.01 & & \\
         &              &               &      &      &      &      &      & \\        
L14      &  +0.025      &  0.92         & 0.03 & 0.88 & 0.10 & 0.92 & 0.03 & 2 \\
         &  $\pm$~0.013 &  $\pm$~ 0.02  &      & $\pm$~ 0.07 & & $\pm$~ 0.02 & &  \\
         &              &               &      &      &      &      &      & \\        
K05      &  -0.024      &  1.07         & -    & 1.03 & -    & 1.07 & - &    1 \\
         &              &               &      &      &      &      &   &\\
\hline 
3D map   &  +0.081      & -             & -    &      &      &  & & 18 \\
         &  $\pm$~0.006 &               &      &      &      & & & \\
\hline \\
\end{tabular}\par\medskip
{\footnotesize PN06: \citet{paunzen}; L14: \citet{lata}; DAML02: \citet{dias02}; M93: \citet{meynet}; PJ94: \citet{phelps}; K05: \citet{keller}; 3D map: \cite{sale}}
\end{flushleft}
\end{table*}

\begin{table*}
\begin{flushleft}
  \caption{Comparison of age estimates determined in this paper 
for ``standard" clusters with earlier published values .}\label{comparage}
\begin{tabular}{|l|r|r|r|r|r|}
\hline
Reference &  $<$log~($t$)--log~ (t(Publ))$>$ & $\sigma$ & $<$log~($t$)--log~(t(Publ))$>$ & $\sigma$ &  Number  \\
          &   \multicolumn{2}{|c|}{($r-i$,H$_{\alpha}$~index)}  &  \multicolumn{2}{|c|}{(H$_{\alpha}$~index, $W$)} & of clusters\\
\hline
PN06     &     +0.30 & 0.40 & +0.36 & 0.45 & 5 \\
         &     $\pm$~ 0.18 & & $\pm$~ 0.20 & &   \\
         &          &      &      &      & \\        
PJ94     &     +0.39 & 0.28 & +0.52 & 0.35 & 13 \\
         &     $\pm$~ 0.08 & & $\pm$~ 0.10 & &       \\
         &          &      &      &      & \\        
M93      &     +0.23 & 0.11 & +0.61 & 0.39 & 3 \\
         &     $\pm$~ 0.07 & & $\pm$~ 0.22 &  \\
         &          &      &      &      & \\        
L14      &     +0.32 & 0.02& +0.21 & 0.13 & 2 \\
         &     $\pm$~ 0.01 & & $\pm$~ 0.09 &   \\
         &            &      &      &      & \\        
K05      &      +0.14 & -    & +0.18 & - &    1 \\
\hline \\
\end{tabular}\par\medskip
{\footnotesize PN06: \citet{paunzen}; L14: \citet{lata}; DAML02: \citet{dias02}; M93: \citet{meynet}; PJ94: \citet{phelps}; K05: \citet{keller}; 3D map: \cite{sale}}
\end{flushleft}
\end{table*}

\begin{table*}
\centering
\caption{H$_{\alpha}$~index offset ($\Delta$~H~$_{\alpha}$-index ) estimates for ``standard" clusters.}\label{deltah} 
\begin{tabular}{lrrrr}
Cluster & $\Delta$H$_{\alpha}$~index & $\Delta$H$_{\alpha}$~index & $\Delta$H$_{\alpha}$~index  & Average per cluster \\
        & ($r-i$,H$_{\alpha}$~index) & (H$_{\alpha}$~index, $r$)  & (H$_{\alpha}$~index, $W$)  & $<\Delta$H$_{\alpha}$~index$>$  \\
\hline
Be 7       & +0.051~$\pm$~0.003      & +0.044~$\pm$~0.002          &  +0.046~$\pm$~0.004 &  {\bf +0.046~$\pm$~0.002} \\
Be 62      & +0.060~$\pm$~0.006      & +0.059~$\pm$~0.003          &  +0.066~$\pm$~0.005 &  {\bf +0.061~$\pm$~0.002}\\
King 12    & +0.081~$\pm$~0.004      & +0.085~$\pm$~0.002          &  +0.075~$\pm$~0.003 &  {\bf +0.082~$\pm$~0.002}\\
NGC 103    & +0.059~$\pm$~0.006      & +0.042~$\pm$~0.003          &  +0.051~$\pm$~0.003 &  {\bf +0.044~$\pm$~0.002}\\
NGC 129    & +0.066~$\pm$~0.008      & +0.055~$\pm$~0.004          &  +0.050~$\pm$~0.003 &  {\bf +0.053~$\pm$~0.002}\\
NGC 436    & +0.060~$\pm$~0.002      & +0.058~$\pm$~0.002          &  +0.053~$\pm$~0.003 &  {\bf +0.058~$\pm$~0.001}\\
NGC 457    & +0.039~$\pm$~0.003      & +0.026~$\pm$~0.002          &  +0.032~$\pm$~0.002 &  {\bf +0.031~$\pm$~0.001}\\
NGC 581    & +0.034~$\pm$~0.003      & +0.022~$\pm$~0.007          &  +0.027~$\pm$~0.005 &  {\bf +0.031~$\pm$~0.002}\\
NGC 654    & +0.042~$\pm$~0.006      & +0.035~$\pm$~0.003          &  +0.035~$\pm$~0.003 &  {\bf +0.036~$\pm$~0.002}\\
NGC 659    & +0.026~$\pm$~0.007      & +0.034~$\pm$~0.003          &  +0.031~$\pm$~0.004 &  {\bf +0.032~$\pm$~0.002}\\
NGC 663    & +0.059~$\pm$~0.007      & +0.059~$\pm$~0.003          &  +0.058~$\pm$~0.002 &  {\bf +0.058~$\pm$~0.002}\\
NGC 869    & +0.047~$\pm$~0.005      & +0.045~$\pm$~0.003          &  +0.042~$\pm$~0.006 &  {\bf +0.045~$\pm$~0.002}\\
NGC 884    & +0.058~$\pm$~0.004      & +0.042~$\pm$~0.002          &  +0.043~$\pm$~0.003 &  {\bf +0.045~$\pm$~0.002}\\
NGC 6834   & +0.084~$\pm$~0.006      & +0.076~$\pm$~0.003          &  +0.077~$\pm$~0.002 &  {\bf +0.077~$\pm$~0.002}\\
NGC 6871   & +0.058~$\pm$~0.003      & +0.059~$\pm$~0.004          &  +0.059~$\pm$~0.003 &  {\bf +0.059~$\pm$~0.002}\\
NGC 7790   & +0.049~$\pm$~0.004      & +0.043~$\pm$~0.004          &  +0.044~$\pm$~0.004 &  {\bf +0.045~$\pm$~0.002}\\
Stock 24   & +0.045~$\pm$~0.004      & +0.056~$\pm$~0.003          &  +0.050~$\pm$~0.004 &  {\bf +0.052~$\pm$~0.002}\\
Trumpler 1 & +0.034~$\pm$~0.025      & +0.049~$\pm$~0.006          &  +0.041~$\pm$~0.006 &  {\bf +0.045~$\pm$~0.004}\\
\hline           
{\bf Mean}       & {\bf +0.053~$\pm$~0.001}      & {\bf +0.051~$\pm$~0.001}          &  {\bf +0.051~$\pm$~0.001} &  {\bf +0.051~$\pm$~0.001}\\
\hline
\end{tabular}
\end{table*}

\section{CONCLUSIONS}

We propose a version of the Q-method adapted  for determining the extinction and distances for young 
open clusters based on H$_{\alpha}$ri photometry from the 
IPHAS survey. The method uses the coeval nature 
of the cluster and the fact that cluster stars are concentrated in a small space and sky area and
is highly robust against age and metallicity variations over a wide range spanning
log~($t$)=6.5--8.5 and [Fe/H]=-0.5~to~+0.5. The colour excesses and 
distances for 18 well-studied clusters  determined using our method agree well with 
the most bona fide published values inferred using the classical technique based
on the analysis of $UBV$ photometric data. However, our age estimates agree rather poorly
with the results published by other authors because of the lack of bright evolved  stars 
in our diagrams due to the saturation of bright-star images in IPHAS frames. 
We found that the observed extinction-free
Q-indices  H$_{\alpha}$~index =  0.755$r$~+~0.245$i$~–-~$H_{\alpha}$ are systematically 
greater than the theoretical values by $\Delta$~H~$_{\alpha}$-index~=~0.051~$\pm$~0.014
and that this offset is practically the same for all the 18 clusters studied and shows 
no trend with reddening. We
interpret this discrepancy as due to some mismatch between the actual and theoretical
zero points of the photometric system. After correction for the inferred 
$\Delta$~H~$_{\alpha}$-index offset the H~$_{\alpha}$-indices of stars with $(r-i)_0>0.1$
and ages in the log~($t$)~=7.0--8.5 interval can be used to determine rather accurate 
individual extinction values and hence to study clusters with variable extinction across
their field. 
 The results obtained
show that our technique is quite well suited for  determination
of open cluster parameters within broad solar neighbourhood.

\section*{ACKNOWLEDGEMENTS}

We thank the anonymous reviewer for the valuable comments, which greatly improved the 
final version of the paper. 
This paper makes use of data obtained as part of the INT Photometric 
H$_{\alpha}$ Survey of the Northern Galactic Plane (IPHAS, www.iphas.org) 
carried out at the Isaac Newton Telescope (INT). The INT is operated 
on the island of La Palma by the Isaac Newton Group in the Spanish 
Observatorio del Roque de los Muchachos of the Instituto de Astrofisica de Canarias. 
All IPHAS data are processed by the Cambridge Astronomical Survey Unit, 
at the Institute of Astronomy in Cambridge. The bandmerged DR2 catalogue was 
assembled at the Centre for Astrophysics Research, University of 
Hertfordshire, supported by STFC grant ST/J001333/1. This research was made possible through the use of the AAVSO Photometric All-Sky Survey (APASS), funded by the Robert Martin Ayers Sciences Fund.
The algorithm and new technique of the determination of fundamental open-cluster 
parameters was supported by the Russian Scientific Foundation 
(grant no. 14-22-00041), and  data mining and determination of the parameters 
of new clusters, by the joint grant by the Russian Foundation for Basic Research and Department of Science and 
Technology of India through project no. RFBR 15-52-45121 and INT/RUS/RFBR/P-219.

%%%%%%%%%%%%%%%%%%%%%%%%%%%%%%%%%%%%%%%

\end{document}